\documentclass[preprint,12pt]{article}
\usepackage{amssymb}
\usepackage{amsmath}
\usepackage{subcaption}
\usepackage{graphicx}
\usepackage{gensymb}
\usepackage{float}
\usepackage{color}

\begin{document}


{\center{\Large{
The influence of structure geometry and material on seismic metamaterial performance}}

\center{\small{T. Venkatesh Varma$^1$, Bogdan Ungureanu$^2$, Saikat Sarkar$^1$, Richard Craster$^{2,3,4}$, S\'ebastien Guenneau$^{4}$,
St\'ephane Br\^ul\'e$^5$}}
\center{\small{$^{1,*}$Discipline of Civil Engineering, Indian Institute of Technology, Indore, India}}
\center{\small{$^2$Department of Mathematics, Imperial College London, London SW7 2AZ, United Kingdom}}
\center{\small{$^3$Department of Mechanical Engineering, Imperial College London, London SW7 2AZ, United Kingdom}}
\center{\small{$^4$UMI 2004 Abraham de Moivre-CNRS, Imperial College London, London SW7 2AZ, United Kingdom}}
\center{\small{$^5$Aix Marseille Univ, CNRS, Centrale Marseille, Institut Fresnel, France}}
\center{\small{$^{*}$Corresponding Author, E-mail: b.ungureanu@imperial.ac.uk}}}


\begin{abstract}
Diverting, and controlling, elastic vibrations impacting upon   infrastructure is a major challenge for seismic hazard mitigation, and for the reduction of machine noise and vehicle vibration in the urban environment. Seismic metamaterials (SMs), with their inherent ability to manipulate wave propagation, provide a key route for overcoming the technological hurdles involved in this challenge. Engineering the structure of the SM serves as a basis to tune and enhance its functionality, and inspired by split rings, swiss-rolls, notch-shaped and labyrinthine designs of elementary cells in electromagnetic and mechanical metamaterials, we investigate altering the structure geometries of SMs with the aim of  creating large bandgaps \textcolor{black}{in a subwavelength regime}. We show that square stiff inclusions, perform better in comparison to circular ones, whilst keeping the same filling fraction. En route to enhancing the bandgap, we have also studied the performance of SMs with different constituent materials; we find that steel columns, as inclusions, show large bandgaps, however, the columns are too large for steel to be a feasible material in practical or financial terms. Non-reinforced concrete would be preferable for industry level scaling up of the technology because, concrete is cost-effective, easy to cast directly at the construction site and easy to provide arbitrary geometry of the structure. As a part of this study, we show that concrete columns can also be designed to exhibit bandgaps if we cast them within a soft soil coating surrounding the protected area for various civil structures like a bridge, building, oil pipelines etc. Although our motivation is for ground vibration, and we use the frequencies, lengthscales and material properties relevant for that application, it is notable that we use the equations of linear elasticity and our investigation is more broadly relevant in solid mechanics.  
  
\end{abstract}





\section{Introduction}
\label{S1}

Controlling elastic waves near structures is key to addressing a large class of civil engineering problems ranging from seismic hazard mitigation to stopping unwanted noise and vibration produced by  vehicles, heavy machinery on construction sites  etc.~\cite{bajcar2012impact,brownjohn2011vibration,zhang2016response,yao2019vibration}. Seismic waves lead to the destruction of civil installations and loss of life, thus forming one of the most important and difficult challenges for civil engineers; the reduction of urban noise and vibration is also important because they create health issues and have other consequences such as affecting highly sensitive scientific and medical instruments. There are many other applications, e.g., in the context of ground motions produced from even minor tremors, industrial machinery or suburban rail transport systems degrading the structural integrity of nearby buildings, pipeline systems, sensitive instruments; additionally, small-scale damage to structures in petrochemical industries or nuclear reactors can have devastating consequences. Traditional approaches for vibration mitigation, mostly in the form of base isolation (BI) and tuned mass damper (TMDs) have certain limitations \cite{zhou2002seismic,xiang2012periodic,xiang2014seismic,harvey2016review,anajafi2018comparison,fabrizio2019tuned}: BI induces large movement of the structures which may not be acceptable, TMD works only for a narrow frequency range outside of which it may work adversely which is again a major issue because identifying the frequency contents in the structure correctly is not straightforward as it may change significantly with time due to structural degradation and environmental change \cite{wagner2017robust,moser2011environmental}. 

\textcolor{black}{In the parallel research area of electromagnetism, there is a keen interest in composites, known as metamaterials, which are assemblies of multiple elements usually arranged in periodic patterns at subwavelength scales, that were introduced \cite{pendry1999}
 as a means to achieve effective electromagnetic properties not from the properties of the constituent materials, but from the combination of elements with given shapes and orientations, such as split ring resonators. Such periodically arranged elements take their name from the shape of the thin metal sheets they are made of that allow for artificial magnetism, not present in the constituent materials (metal surrounded by plastic, neither of which are magnetic media), through the precise manipulation of electric and magnetic components of electromagnetic field and their interplay. Indeed, the electromagnetic field is tremendously enhanced due to internal capacitance and inductance phenomena upon resonance of the thin sheets.
 From the highly dispersive nature of the metamaterials around the resonant frequencies of split ring resonators and other arrangements such as swiss-rolls, it is possible to block, absorb, enhance or even bend electromagnetic waves propagating through a doubly or triply periodic array of them, as low frequency stop bands and strong anisotropy (apparent in distorted isofrequency contours in pass bands) take place \cite{kadic2013}.}

Metamaterials have the capability of manipulating a desired range of frequency components in the propagating wave, an aspect that has been extensively used in electromagnetics, optics and micro and nano-scale mechanics \cite{watts2012metamaterial,casablanca2018seismic,craster2012acoustic,guenneau2007acoustic,poddubny2013hyperbolic,kadic2019static,brule2019role,zeng2020matryoshka,brule2018seismic,brule2017metamaterial,colombi2017elastic,yang2016metasurface}. However, their large scale extension, which can be extremely useful in solving many of the important engineering problems, mentioned above, requires development. Recent developments have indicated that metamaterials, if designed appropriately, can indeed provide a robust solution for elastic waves, thanks to their ability to create large frequency bandgaps even at large scale \cite{palermo2016engineered,liu2019trees,brule2014experiments,ungureanu2019influence,colombi2016seismic,moitra2015large,aravantinos2015large,mosby2016computational,ungureanu2015auxetic,diatta2016control,sun2019bandgap,muhammad2019seismic,del2014dynamic}.

Early work by Economou and Sigalas \cite{economou1993classical} established a generic trend that a denser inclusion in the microstructure geometry of periodic media exhibits bandgaps for 2D and 3D structures. This phenomenon is predicted in \cite{mandelik2003band,gazalet2013tutorial,gomez2013floquet} by using the Floquet-Bloch theory which is applicable to different wave types traveling through a periodic media. Experimentally \cite{meseguer1999rayleigh,meseguer1999two} showed the attenuation of surface elastic waves in a marble quarry containing repeated circular holes with the bandgaps obtained in a high frequency range of several  kHz which is not important for seismic applications, because seismic forces mostly contain very low frequencies in the range of few Hz  \cite{gadallah2005applied,kramer2007geotechnical,chestler2017evidence,nakano2018shallow}. The first full-scale experiments \cite{brule2014experiments,brule2017flat} to attenuate surface elastic waves, such as Rayleigh and Love waves, were conducted in structured soil and extended in  \cite{ungureanu2019influence,brule2020emergence,brule2019roleb,brule2018experimental}; the soil was engineered with cylindrically configured voids as inclusions in a periodic manner and this experiment shows the feasibility of using phononic crystals and metamaterials at a meter-scale important for civil engineering applications. 
 
A variety of extensions of these concepts are aimed at creating larger bandgaps and to force these to occur at the low frequencies required for civil engineering applications. Recently \cite{miniaci2016large} numerically analyzed and designed optimal configuration for inclusions of micro-structure geometry, i.e., cross-shaped voids, hollow cylindrical and locally resonant inclusions (e.g. steel, rubber, concrete), which shield low frequency contents in seismic forces. Their parametric study on the filling fraction of inclusions in micro-structure shows a possibility of enhancing the bandgap. The ultimate goal of bandgap engineering in this context is to create a zero-frequency bandgap, i.e. one that starts at zero frequency and then extends over a broad range of low frequencies. \cite{achaoui2017clamped} show that this can be achieved with cylindrical steel inclusions clamped to a bed rock which lies underneath a soil layer; they also considered struts linking the cylindrical steel columns.

\begin{figure}[h!]
\begin{center}
\includegraphics[width=10cm]{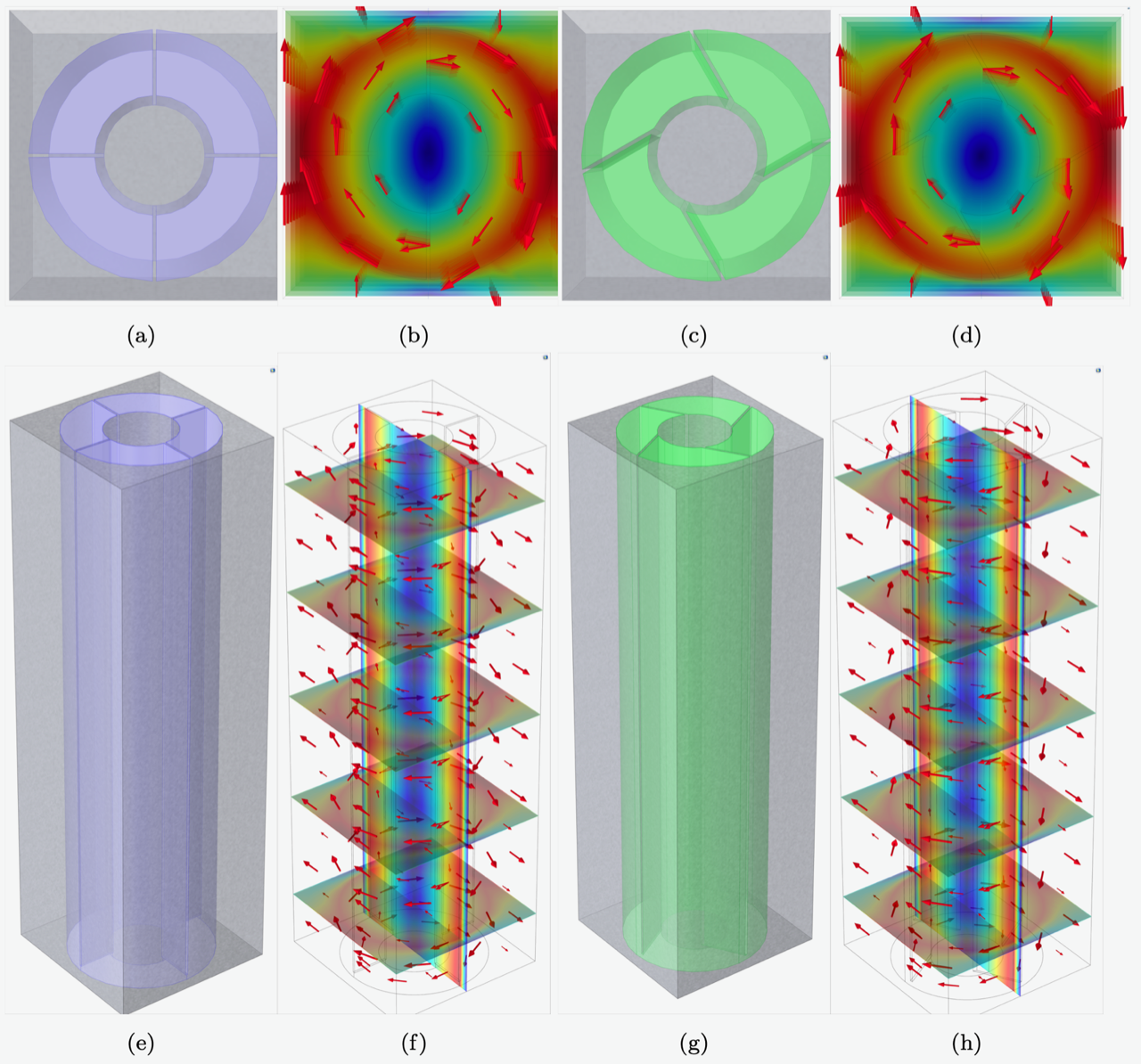}
\end{center}
\caption{Unclamped cylindrical steel inclusion (radius $r=0.62$ m, height $h=10$ m) placed inside a cylindrical air cavity ($r=1.3$ m, $h=10$ m), surrounded by soil, inside an elementary cell ($3$ m $\times 3$ m $\times 10$ m) with periodic boundary conditions on opposite vertical sides, thus structuring soil in a doubly periodic manner. Each steel cylinder is connected to the soil matrix with 4 ligaments ($0.04$ m thick steel plates, $10$ m in height). When ligaments are aligned with the center of the cell (a,e), a rotational eigenmode (b,d) occurs at frequency 42 Hz.  When ligaments are tilted by an angle of $30$ degrees with respect to the center of the cell (c,g), the rotational eigenmode (d,h) is shifted to frequency 38.6 Hz.}\label{F0}
\end{figure}

Here we focus on large scale metamaterials made by periodically installing stiff columns in a matrix-like soil. There are many aspects that are not yet known about such metamaterials, e.g., effect of structure geometry, that is the detailed geometry within a single building block of the periodic medium, and material constituents. 
Although the possibility of enhancing the bandgap via varying the substitution ratio of stiffer inclusions is established, the effect of varying the structure geometry, keeping the same filling-fraction, is yet to be addressed. In the present study, we investigate the effect of the structure geometry in SMs with the aim of creating large bandgaps, keeping volume of inclusions same. As an illustrative example of the importance of structure geometry, we show in Figure \ref{F0} two similar geometries for SMs: we see in Figure A.18 that just simply tilting some junctions between a cylinder and a bulk medium has a profound impact on stop band width, as noted in earlier studies on phononic crystals consisting of inertial resonators \cite{bigoni2013,wang2014,wang2015}. We also explore the performance of SMs by considering different constituent materials. Specifically, we aim to maximize the bandgap width as well as the number of stopbands since the SMs may have to safeguard different types of civil installations with varied frequency contents ranging from massive tall buildings with very small natural frequencies to fluid conveyors containing large natural frequencies \cite{dym2007estimating,dym2012approximating}. En-route to enhancing the bandgap, we report an innovative design for SM, which is very pertinent to civil engineering practices. The rest of the paper is organized as follows: The mathematical formulation is shown in section \ref{S2}  and consequently numerically simulated results are given in section \ref{S3}.  Finally, we draw some conclusions in Section \ref{S4}.

\section{Mathematical Formulation-Finite Element Approach}         
\label{S2}

We take an elastic medium consisting of isotropic homogeneous phases, for which the governing equation \cite{achenbach73a} is 
\begin{equation} \label{e2bis}
(\lambda+\mu)\frac{\partial^2 u_j}{\partial x_j \partial x_i}+\mu\frac{\partial^2 u_i}{\partial x_j \partial x_j}=\rho\frac{\partial^2{u}_i}{\partial t^2}
\end{equation}
which is valid in each homogeneous phase, we use tensor notation. Eq. \eqref{e2bis} is supplied with a no slip condition at the interface between the homogeneous phases, i.e., relative displacement will be zero between the phases as well as continuity of the normal component of stress to the interface. We work primarily in the frequency, $\omega$, domain and assume $\exp (-i\omega t)$ dependence is considered understood and suppressed henceforth. 

A periodic medium can be analyzed by sequential translational operations performed on its elementary cell, by making use of a lattice vector ${\bf a}=(a_1,a_2)$. The inherent periodicity of this cell enables us to characterize the dispersion properties of elastic waves propagating within such a periodic medium, via Bloch's theorem as in Eq.(\ref{e1}), where $\mathbf{k}=\{k_x,k_y\}$ is the Bloch wavevector and $\mathbf{x} =(x_1,x_2)$ \cite{brillouin2003wave}:
\begin{equation} \label{e1}
\mathbf{u(x+\mathbf{a})}=\mathbf{ u(\mathbf{x})}\exp({i\mathbf{k}.\mathbf{a}})
\end{equation}
 and, to obtain, dispersion relations between frequency and phase-shift we need only consider the elementary cell.

For our numerical simulations we arrive at  
Eq. \eqref{e3} by discretizing ${u}$ in weak form using 3D shape functions in a finite element approach \cite{reddy1993introduction}:  $\mathbf{K}$ and $\mathbf{M}$ being global stiffness and mass matrices, respectively, which are basically functions of Bloch wavevector $\mathbf{k}$, and $\mathbf{U}$ is the assembled displacement vector.
\begin{equation} 
\label{e3} (\mathbf{K}-{\omega^2}\mathbf{M})\mathbf{U}=0
\end{equation}


Floquet-Bloch analysis is conducted by computing dispersion curves via varying $\mathbf{k}=\{k_x,k_y\}$ along the edges of irreducible Brillouin zone (IBZ). For a square lattice, IBZ lies along the edges of the triangle $\Gamma M X$; $\Gamma = (0,0)$, $M= (\frac{\pi}{a},\frac{\pi}{a})$ and $X=(\frac{\pi}{a},0)$, where $a=a_1=a_2$. For a real valued $\mathbf{k}$, the frequencies ($\omega$) are obtained by solving an appropriate eigenvalue problem, thereby constructing the dispersion curves via plotting real valued frequencies corresponding to  $\mathbf{k}$ \cite{van2007computation,jarlebring2017waveguide}. Since dispersion curves are computed over a medium representing an infinite array of micro-structured geometry, they alone do not fully reveal the inherent effectiveness of SM design. For this purpose, transmission spectra are also computed, which is a measure of wave propagation attenuation/transmission losses over a finite medium \cite{lee2010transmittance}.    


\section{Results and Discussion}
\label{S3}
In this section, we present dispersion curves and transmission losses for different microstructure geometries. To arrive at a microstructure that maximizes the bandgap, we compare different results keeping the substitution ratio the same. Having arrived at a particular configuration, we explore the effect of changes in orientation of the inclusion from $0\degree$ to $45\degree$. After this we move to the aspect of material constituent for the inclusion and we also focus on cost effectiveness and ease of fabrication of arbitrary microstructure geometry.

\begin{figure}
	\begin{subfigure}[b]{0.3\linewidth}
		\centering
		\includegraphics[width=0.8\linewidth]{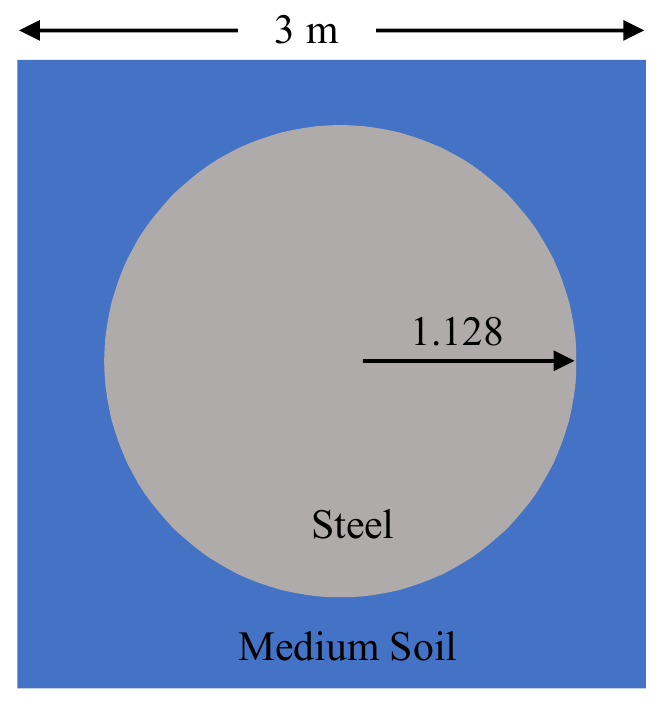} 
		\caption{Cross-section of cylindrical steel inclusion with 0.445 substitution ratio} 
		\label{F1a} 
		\vspace{3ex}
	\end{subfigure}
	\begin{subfigure}[b]{0.7\linewidth}
		\centering
		\includegraphics[width=1.0\linewidth]{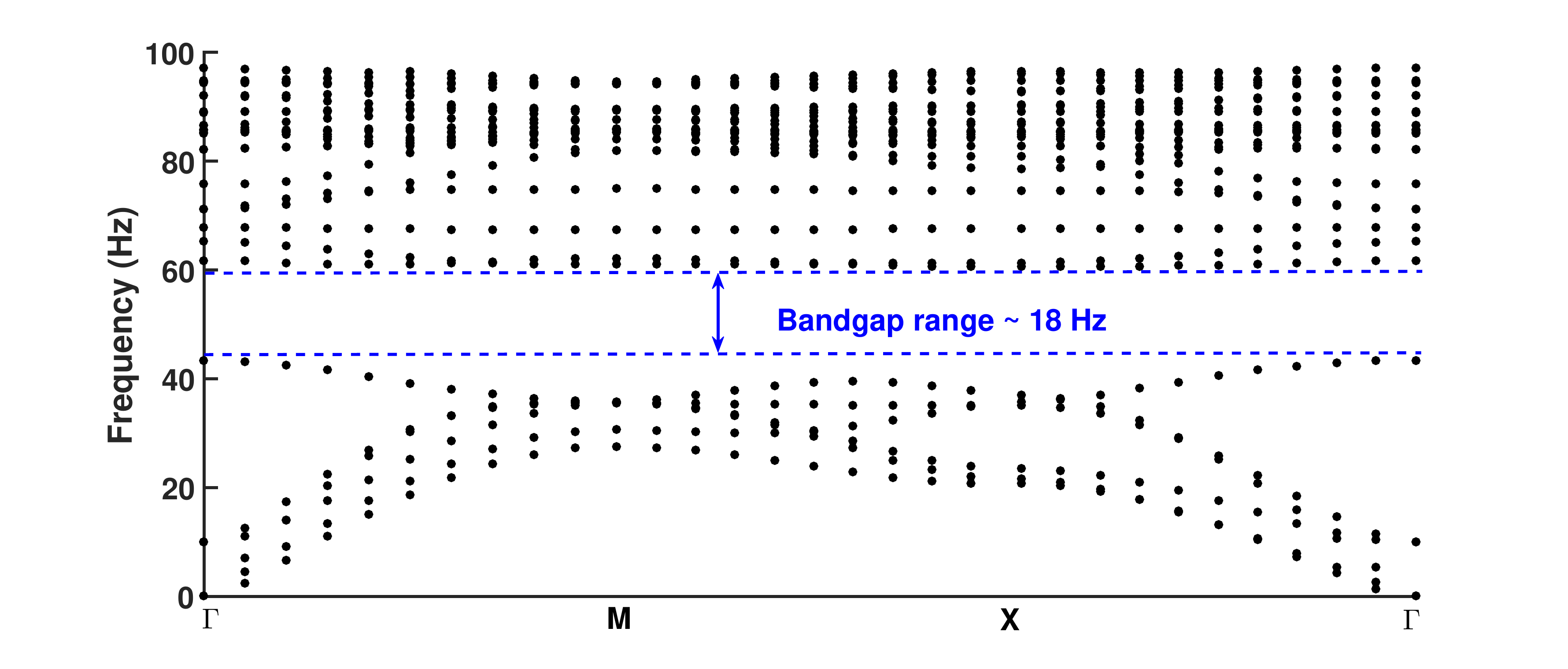} 
		\caption{Dispersion curves for cylindrical inclusion} 
		\label{F1b} 
		\vspace{4ex}
	\end{subfigure} 
	\begin{subfigure}[b]{0.3\linewidth}
		\centering
		\includegraphics[width=0.8\linewidth]{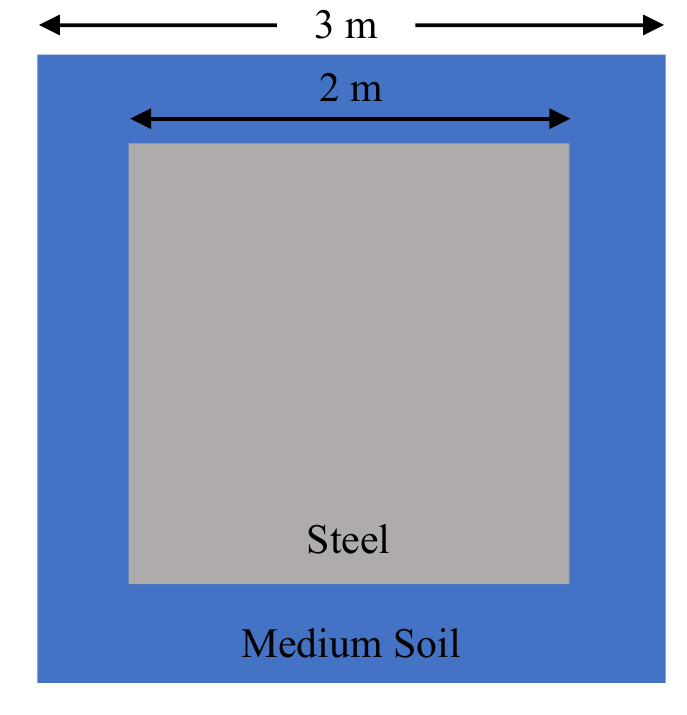} 
		\caption{Cross-section of regular square steel inclusion with 0.445 substitution ratio} 
		\label{F1c} 
		\vspace{-1ex}
	\end{subfigure}
	\begin{subfigure}[b]{0.7\linewidth}
		\centering
		\includegraphics[width=1\linewidth]{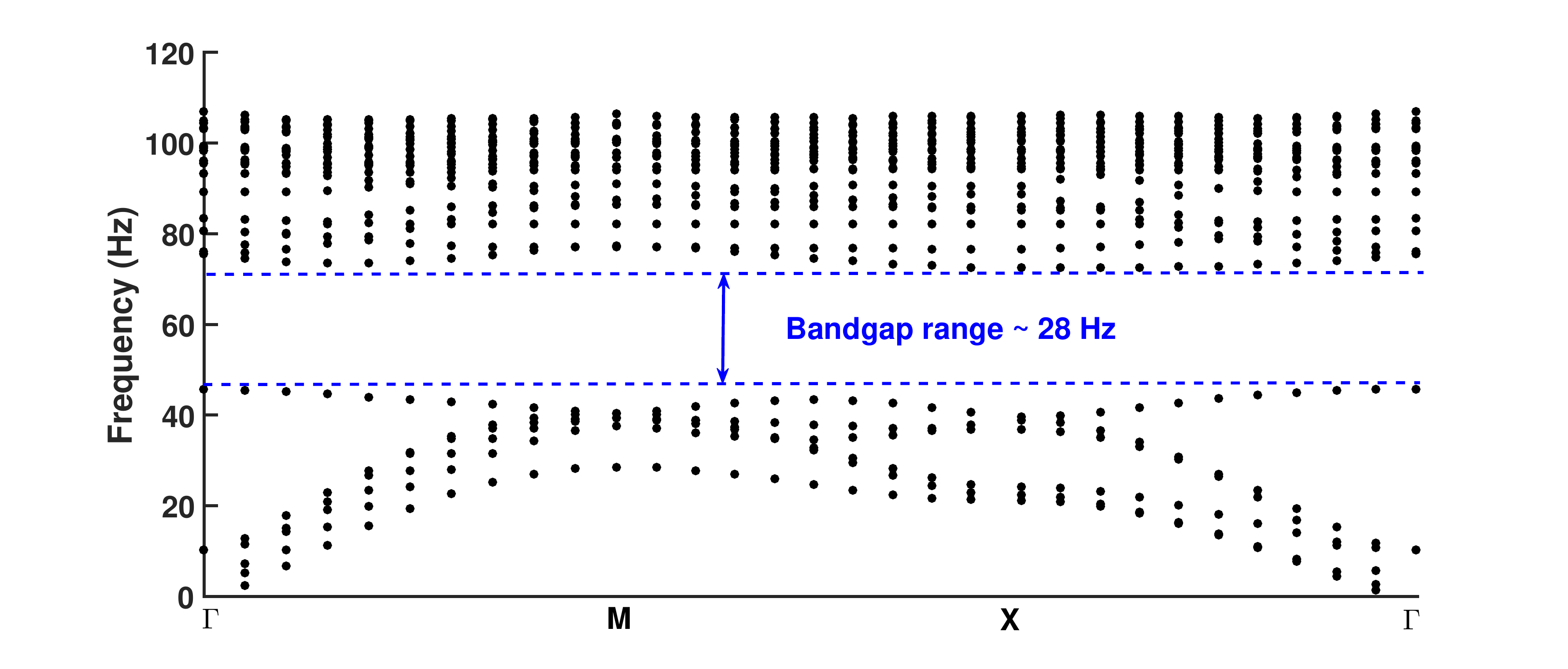} 
		\caption{Dispersion curves for square inclusion} 
		\label{F1d} 
	\end{subfigure}
	\begin{subfigure}[b]{0.3\linewidth}
		\centering
		\includegraphics[width=0.8\linewidth]{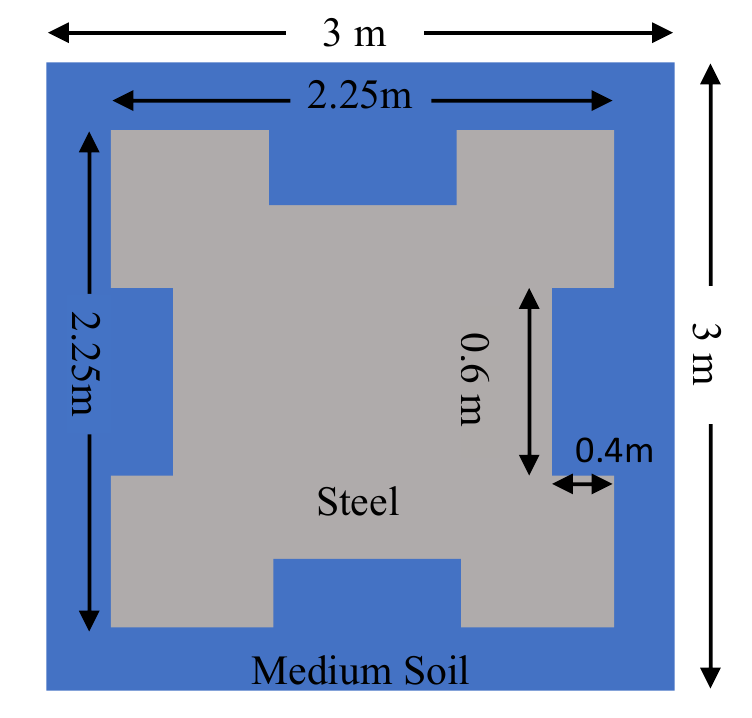} 
		\caption{Cross-section of notch-shaped square steel inclusion with 0.445 substitution ratio} 
		\label{F1e} 
		\vspace{-1ex}
	\end{subfigure}
	\begin{subfigure}[b]{0.7\linewidth}
		\centering
		\includegraphics[width=1\linewidth]{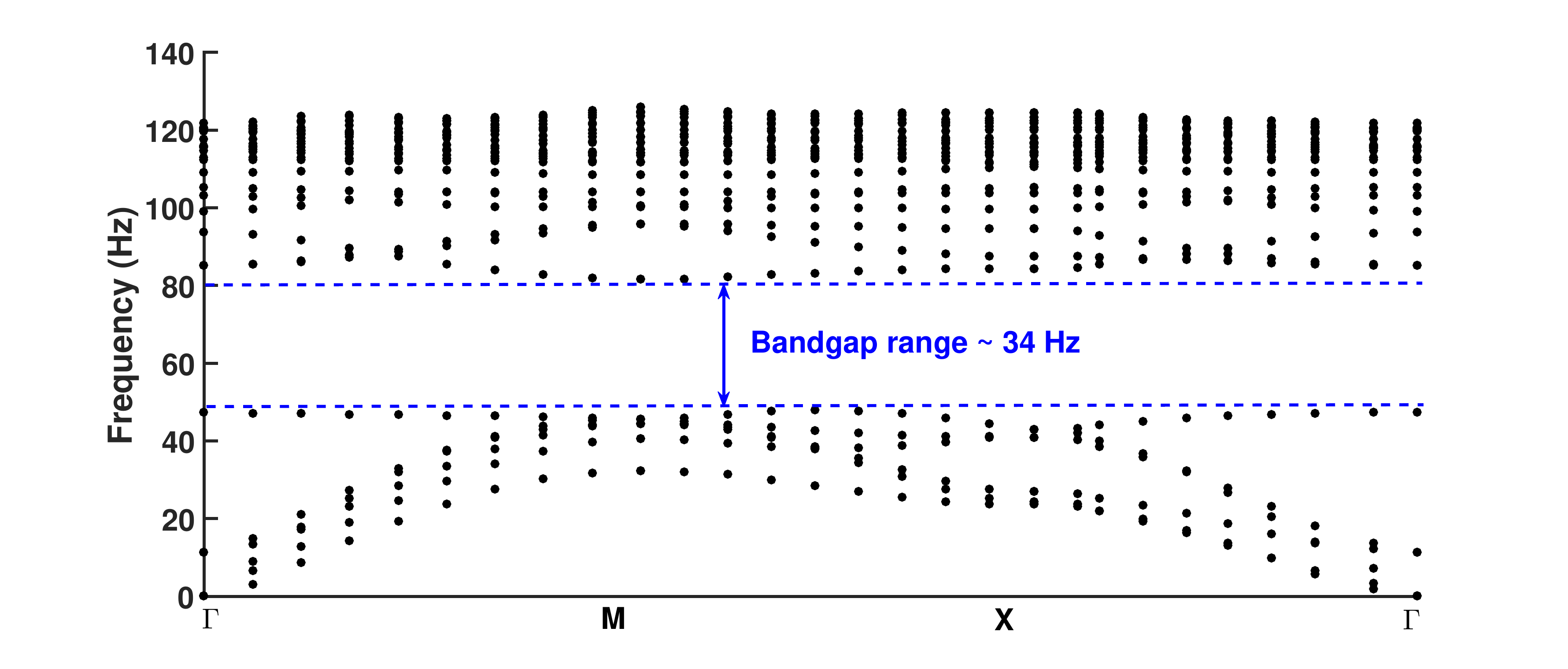} 
		\caption{Dispersion curves for notch-shaped square inclusion} 
		\label{F1f} 
	\end{subfigure}
	\caption{Unclamped configuration of (a) cylindrical ($r=1.128$ m, $h=10$ m), (c) regular-shaped square (2m x2m x10m) and (e) notch-shaped square geometry inclusions in the elementary cell (3m x3m x10m) of periodic media (medium soil) with equal substitution ratio (0.445). Dispersion curves for (b) cylindrical (d) regular-shaped square and (f) notch-shaped square inclusions are obtained around the edges of the irreducible Brillouin zone $\Gamma X M$. Note that there is an enhanced bandgap range with regular-shaped square (approx. 28 Hz) and notch-shaped square  (approx. 34 Hz) inclusions  in comparison to cylindrical inclusion (approx. 18 Hz).}
	\label{F1} 
\end{figure}

\subsection{Comparison of different geometry of inclusions} 
\label{S31}
Most standard civil engineering infrastructures are constructed on medium to dense soil,  for their obvious stability purposes, and to reflect this we choose realistic soil conditions, i.e., medium soil with parameters as elastic modulus, $\text{E}=153 \ \text{MPa}$, Poisson's ratio, $\mu=0.3$ and density, $\rho=1800 \  \text{kg}/\text{m}^3$. The micro-structure of the periodic media is configured with steel columns as inclusions in three forms having same substitution ratio  as shown in Fig. \ref{F1}.
\begin{itemize}
	\item Cylindrical steel column of height 10 m and circular cross-section of radius 1.128 m.
	\item Regular sized square steel column of same height, i.e., 10 m. The cross-section of regular sized square is taken as $2 \ \text{m} \times 2 \ \text{m}$.
	\item Notch-shaped square steel column again with same height, i.e, 10 m. $0.4 \ \text{m}\times 0.6 \  \text{m}$ notch is provided in a square of 2.25 m size as shown in Fig. \ref{F1e}.      
\end{itemize} 

Each form of inclusion is embedded in unstructured medium soil matrix ($3 \ \text{m}\times3 \ \text{m}\times10 \ \text{m}$) so as to make an elementary cell that forms the microstructure of the periodic media; top and bottom of the microstructure remains unclamped for the simulations shown in Fig.\ref{F1}. Elastic properties of steel are chosen as $E=200 \ \text{GPa}$, Poisson's ratio, $\mu=0.33$ and density, $\rho=7850 \  \text{kg}/\text{m}^3$. We have also used soft clay models (which will be seen subsequently) to study the effect of soil on bandgap enhancement.   A comparison of its dispersion properties, obtained along the edges of the irreducible Brillouin zone $\Gamma M X$ (see section \ref{S2}),  can be made from Fig.\ref{F1}. Bandgaps are obtained in all the configurations, i.e., microstructure  with inclusions having circular, regular square and notch-shaped square cross-sections. The square cross-section inclusions show a higher range of bandgap (approximately 28 Hz and 34 Hz in regular and notch-shaped, respectively) in comparison to circular cross-section inclusion (approximately 18 Hz) for the same substitution ratio of steel.

\begin{figure}
	\begin{subfigure}[b]{1\linewidth}
		\centering
		\includegraphics[width=0.8\linewidth]{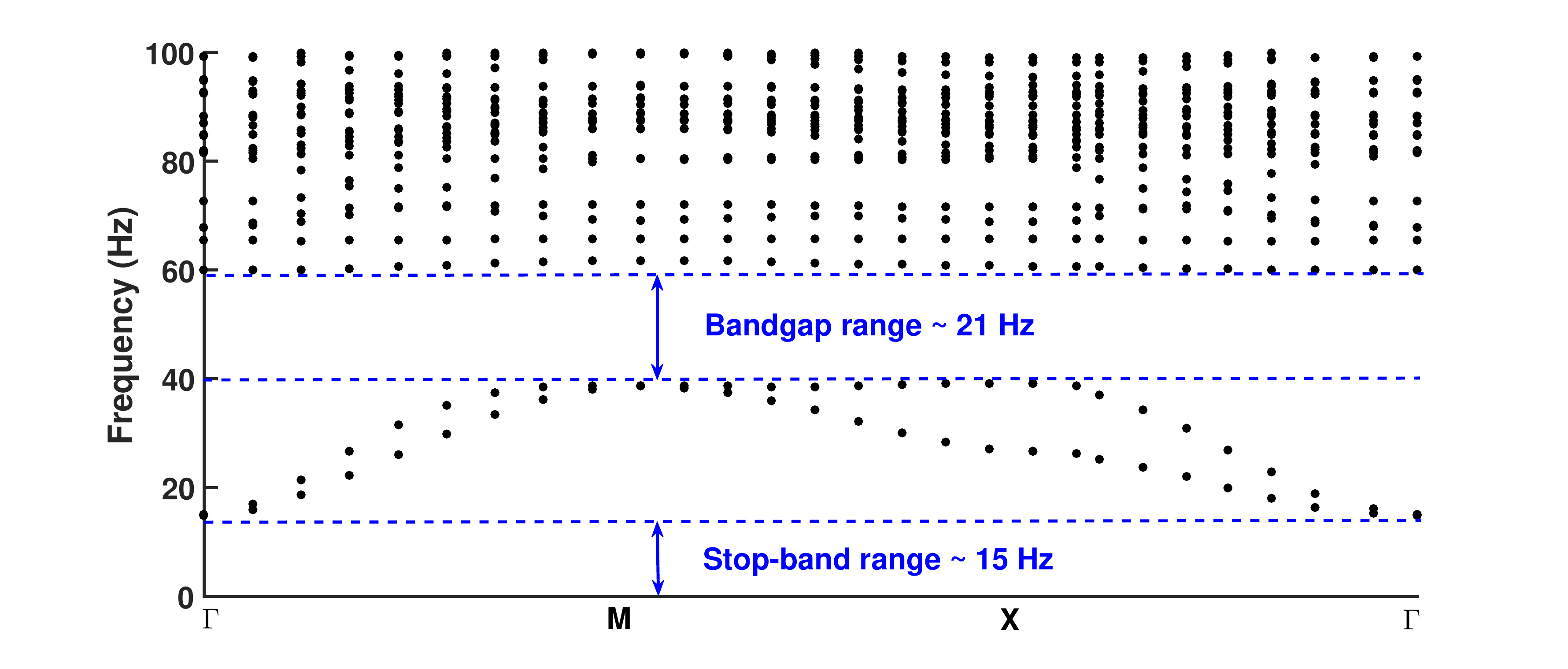} 
		\caption{Dispersion curves for circular cross-section with 1.128 m radius as shown in Fig. \ref{F1a} }
		\label{F9a} 
	\end{subfigure} 
	\begin{subfigure}[b]{1\linewidth}
		\centering
		\includegraphics[width=0.8\linewidth]{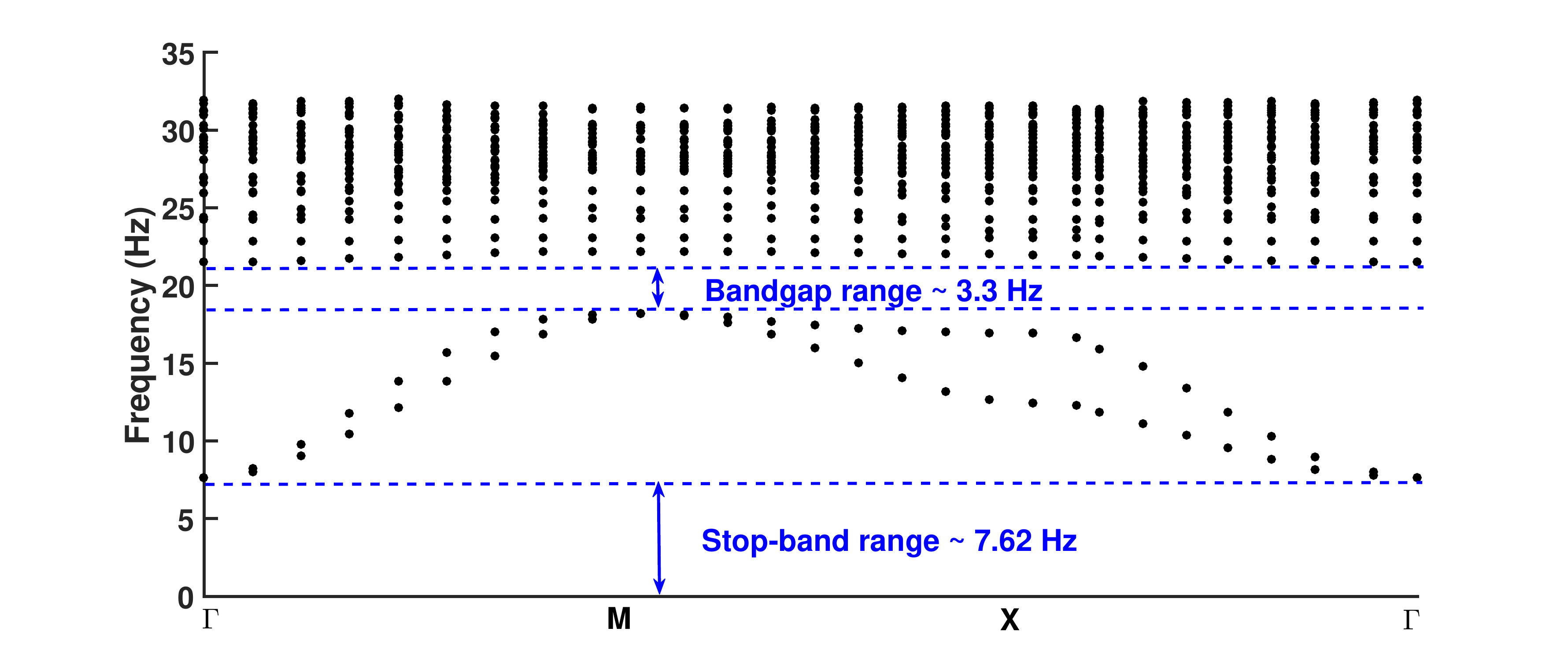} 
		\caption{Dispersion curves for regular-shaped square cross-section of side length 2 m as shown in Fig. \ref{F1c}}
		\label{F9b} 
	\end{subfigure} 
	\begin{subfigure}[b]{1\linewidth}
		\centering
		\includegraphics[width=0.8\linewidth]{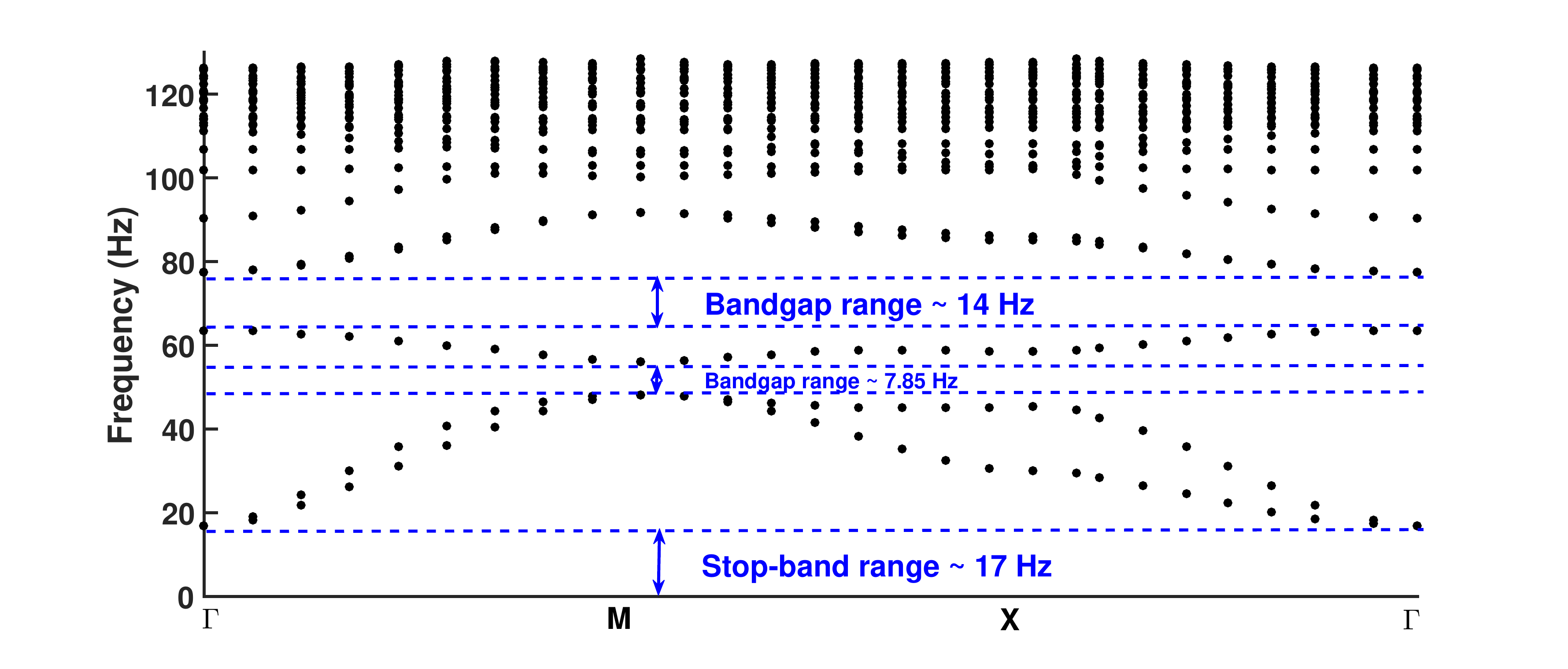} 
		\caption{Dispersion curves for notch-shaped square cross-section (four notches of dimension 0.6 m $\times$ 0.4 m in a 2 m square as shown in Fig. \ref{F1e})} 
		\label{F9c} 
	\end{subfigure} 
	\caption{Dispersion curves for clamped configuration of geometries, i.e. the base of the inclusion column is clamped, shown in Fig.\ref{F1} with the same substitution ratio 0.445. Notch-shaped square cross-section shows a higher stop-band than the other two geometries, however, the frequencies are lifted up in their comparison.}
	\label{F9} 
\end{figure}

Comparison of stop-bands for these configurations are also obtained (see Fig.\ref{F9}), by clamping the bottom of the microstructure. Here, the column inclusions with circular cross-sections show better zero-frequency stop-bands (with upper limit at approximately 15 Hz) in contrast to regular square shaped inclusions (upper limit at approximately 7.62 Hz). However, the notch-shaped square inclusion shows a slightly higher stop-band (upper limit approximately 17 Hz) to that of circular cross-section for same substitution ratio of steel inclusion. Along with the zero-frequency stop-band, there is an additional bandgap at a higher frequency range that is also observed for all three forms of inclusions. The circular cross-section shows a higher frequency bandgap with a range of around 21 Hz in comparison to regular square shaped cross section which has only a narrow bandgap with a range of about 3.3 Hz. The notch-shaped square cross-section shows a slightly higher range of bandgaps, approximately 21.75 Hz (7.85 Hz + 14 Hz), than the circular cross-sectioned inclusion. However, the bandgaps for  circular and notch-shaped inclusions are positioned in the higher frequency spectrum, which are less important for engineering applications.

\begin{figure}[H] 
	\begin{subfigure}[b]{0.3\linewidth}
		\centering
		\includegraphics[width=0.8\linewidth]{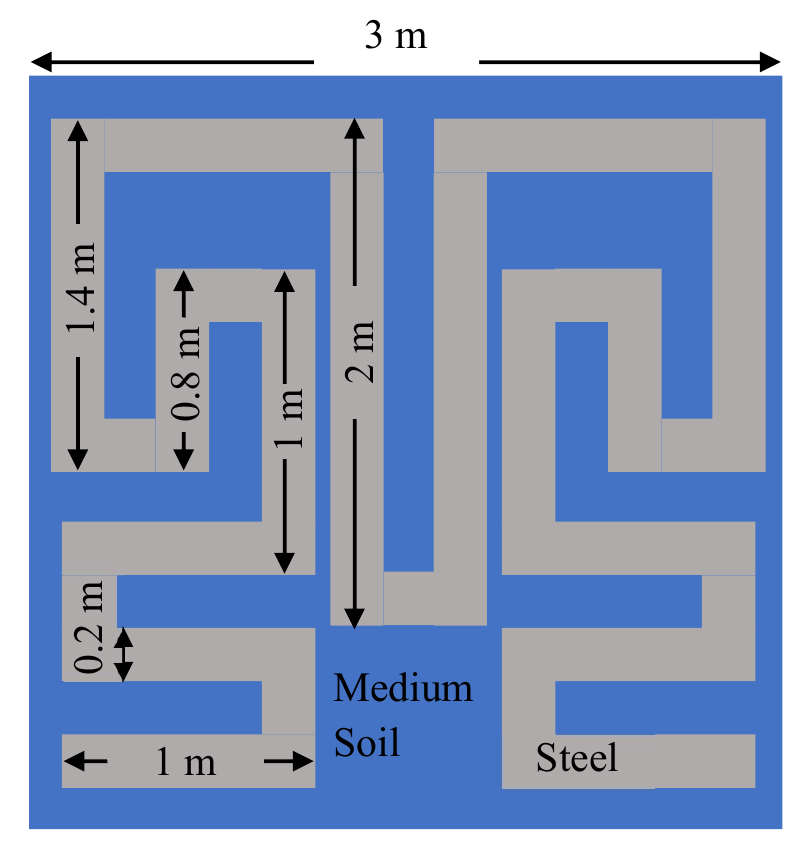}
		\caption{Cross-section of cylindrical coil/Labyrinthine inclusion with substitution ratio 0.445} 
		\label{F11a} 
		\vspace{2ex}
	\end{subfigure}
	\begin{subfigure}[b]{0.7\linewidth}
		\centering
		\includegraphics[width=1.0\linewidth]{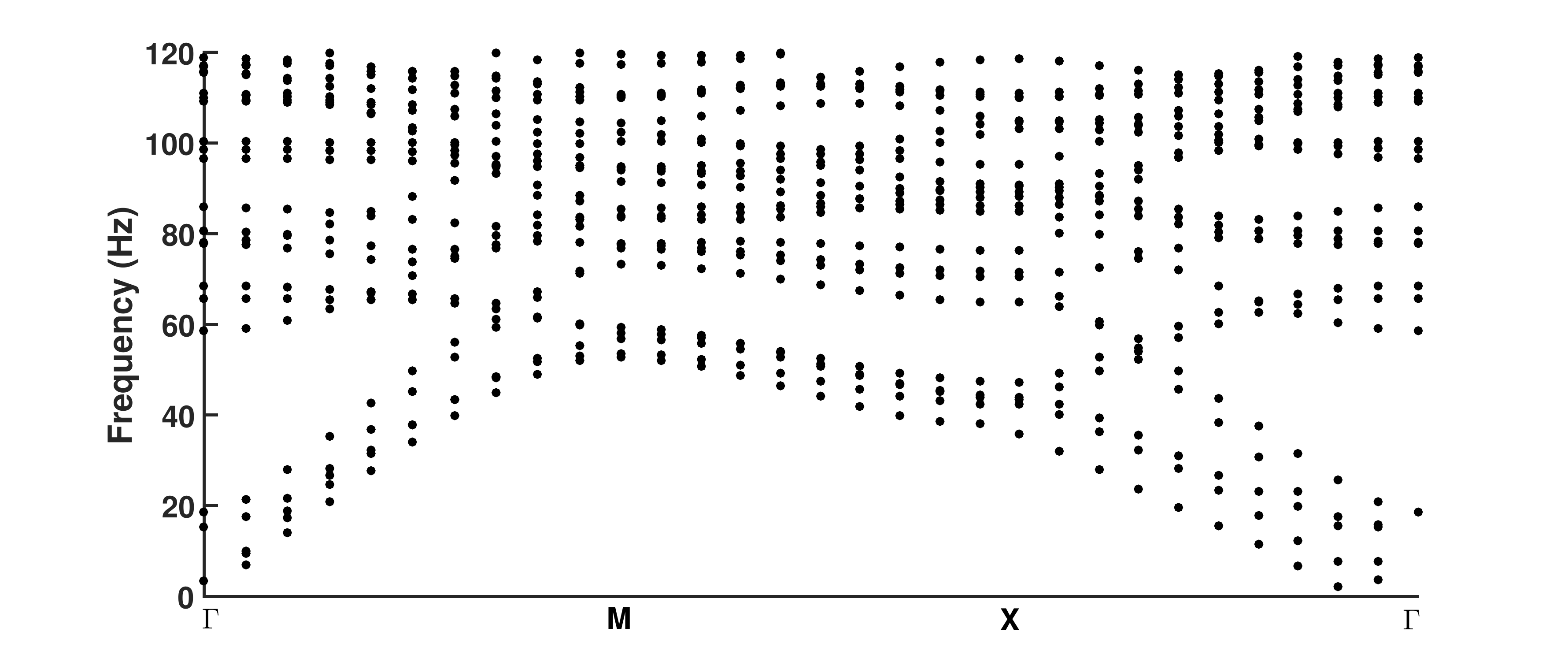} 
		\caption{Dispersion curves for cylindrical coil/Labyrinthine inclusion} 
		\label{F11b} 
		\vspace{2ex}
	\end{subfigure} 
	\begin{subfigure}[b]{0.3\linewidth}
		\centering
		\includegraphics[width=0.8\linewidth]{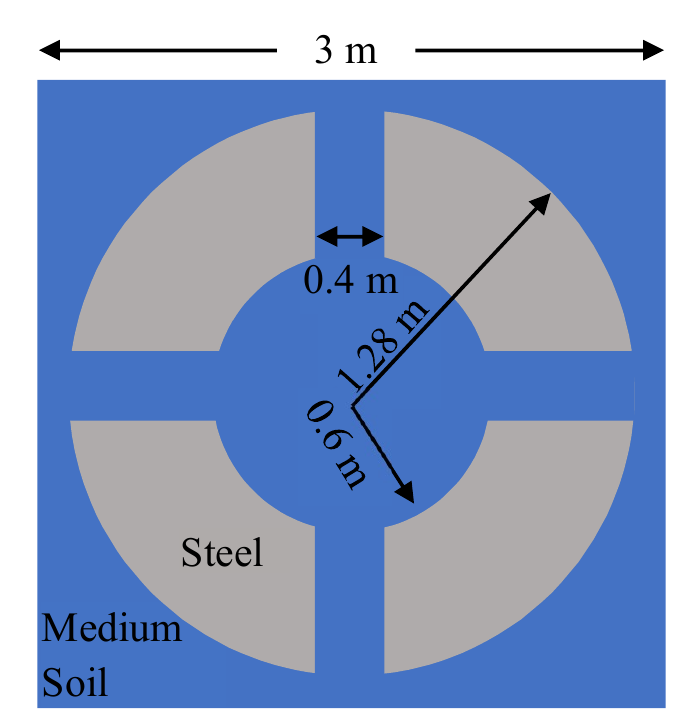} 
		\caption{Cross-section of 4 gaps cylindrical split-ring inclusion with substitution ratio 0.445} 
		\label{F11c} 
		\vspace{-0ex}
	\end{subfigure}
	\begin{subfigure}[b]{0.7\linewidth}
		\centering
		\includegraphics[width=1\linewidth]{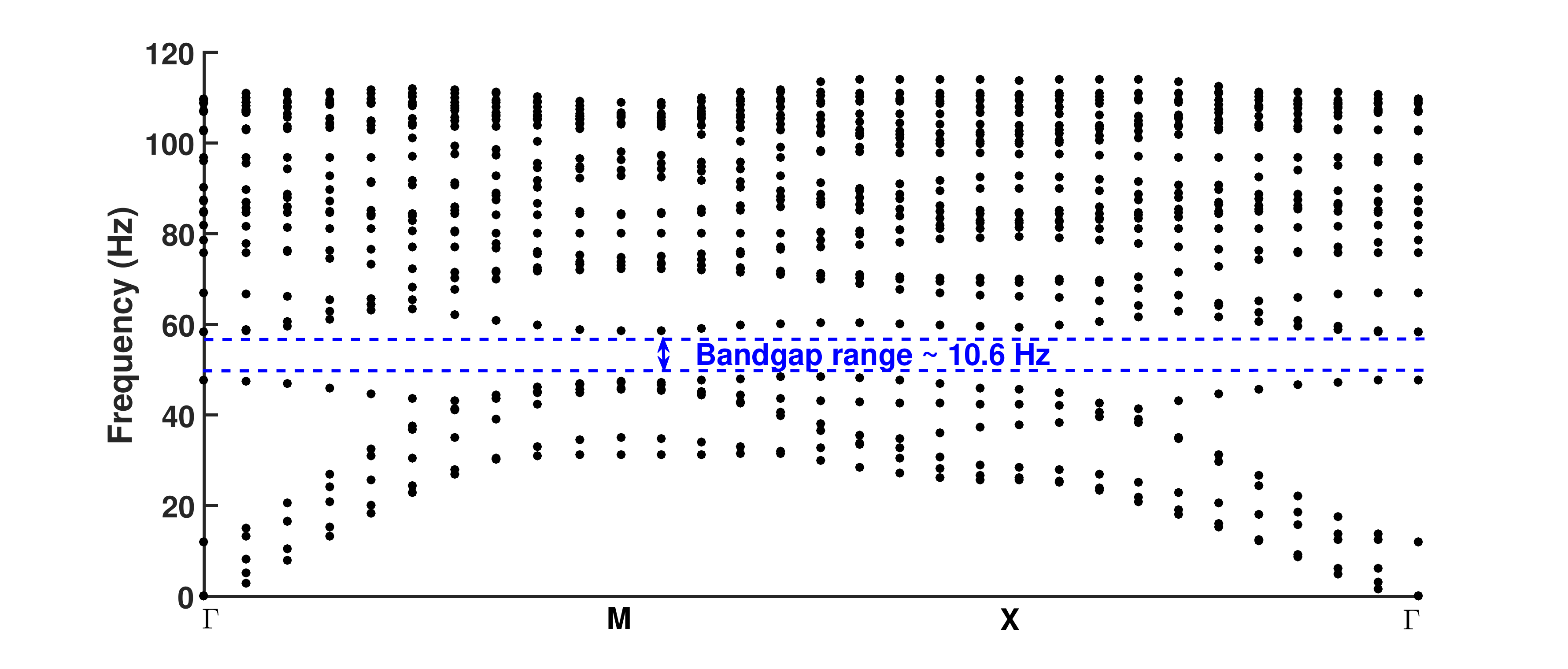} 
		\caption{Dispersion curves for 4 gaps cylindrical split-ring inclusion  } 
		\label{F11d} 
	\end{subfigure}
	\begin{subfigure}[b]{0.3\linewidth}
		\centering
		\includegraphics[width=0.8\linewidth]{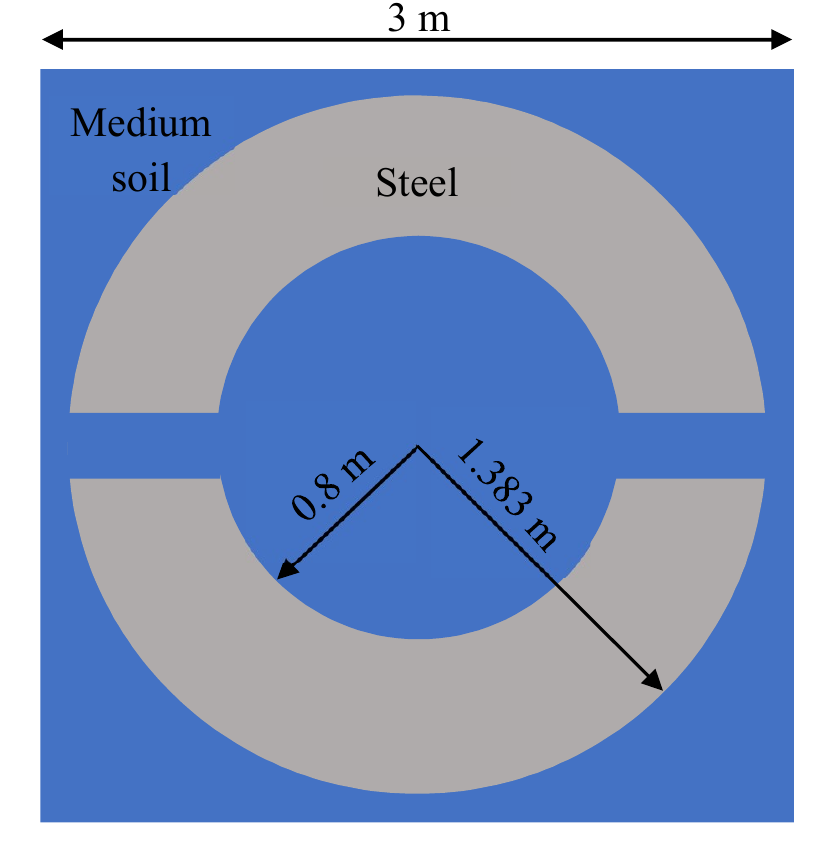} 
		\caption{Cross-section of 2 gaps cylindrical split-ring  inclusion  with substitution ratio 0.445} 
		\label{F11e} 
		\vspace{-1ex}
	\end{subfigure}
	\begin{subfigure}[b]{0.7\linewidth}
		\centering
		\includegraphics[width=1\linewidth]{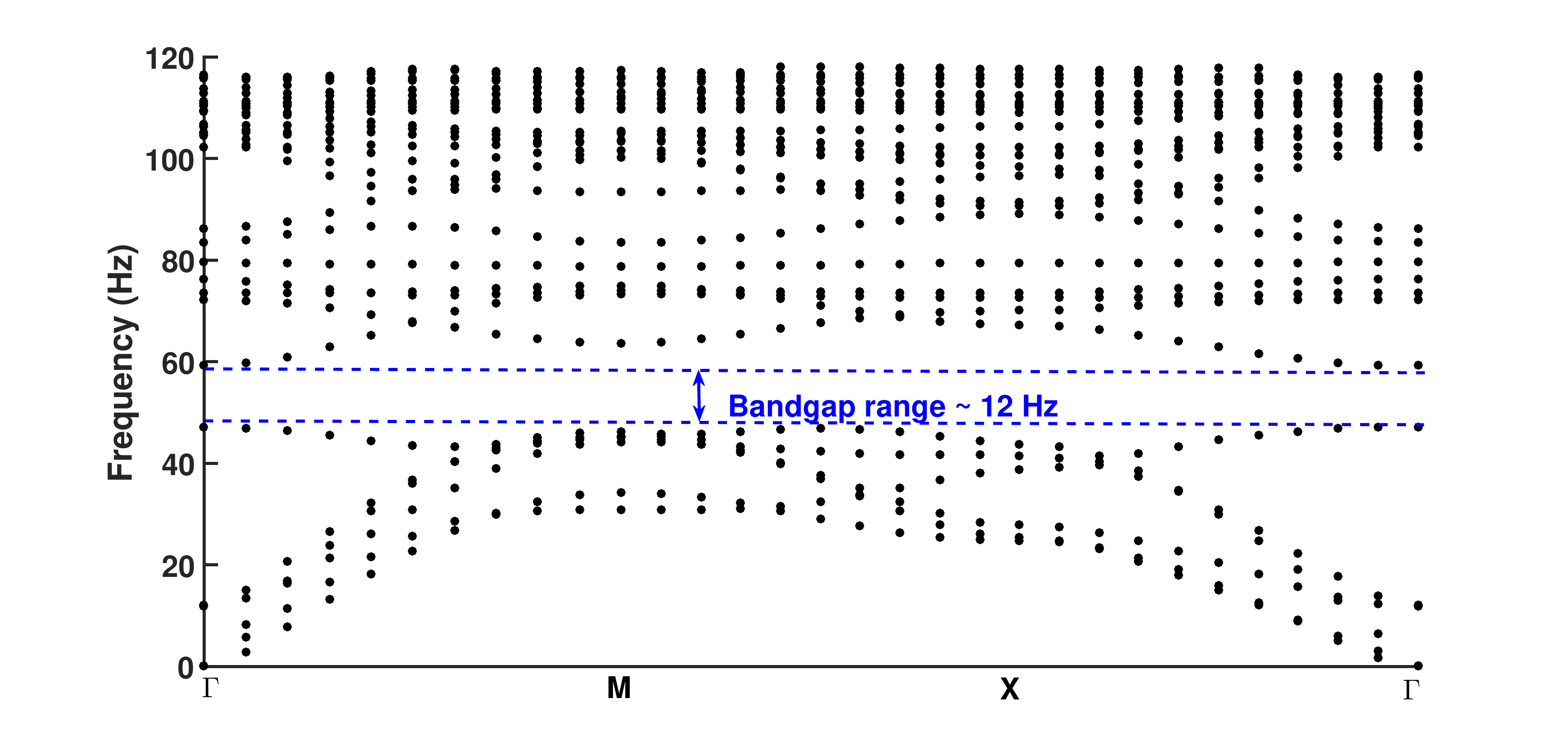} 
		\caption{Dispersion curves for 2 gaps cylindrical split-ring inclusion} 
		\label{F11f} 
	\end{subfigure}
	\caption{Unclamped configuration of (a) coil/Labyrithine, (c) 4 gaps split-ring and (e) 2 gaps split-ring with steel as inclusion with substitution ratio 0.445 in the elementary cell (3m x3m x10m) of periodic media (medium soil).  Corresponding dispersion curves obtained around the edges of the irreducible Brillouin zone $\Gamma X M$ are shown aside (b, d, f). Labyrithine inclusion shows no bandgap in medium soil, whereas split-ring with same substitution ratio shows bandgap in the range of 45 Hz - 60 Hz. Notably, split-ring with 2 gaps shows an enhanced bandgap (12 Hz in the range of 46.84 Hz- 59.23 Hz) as compared to 4 gap split-ring (10.6 Hz in the range of 48.53 Hz- 58.35 Hz).}
	\label{F111} 
\end{figure}

\begin{figure}[H] 
	\begin{subfigure}[b]{0.3\linewidth}
		\centering
		\includegraphics[width=0.8\linewidth]{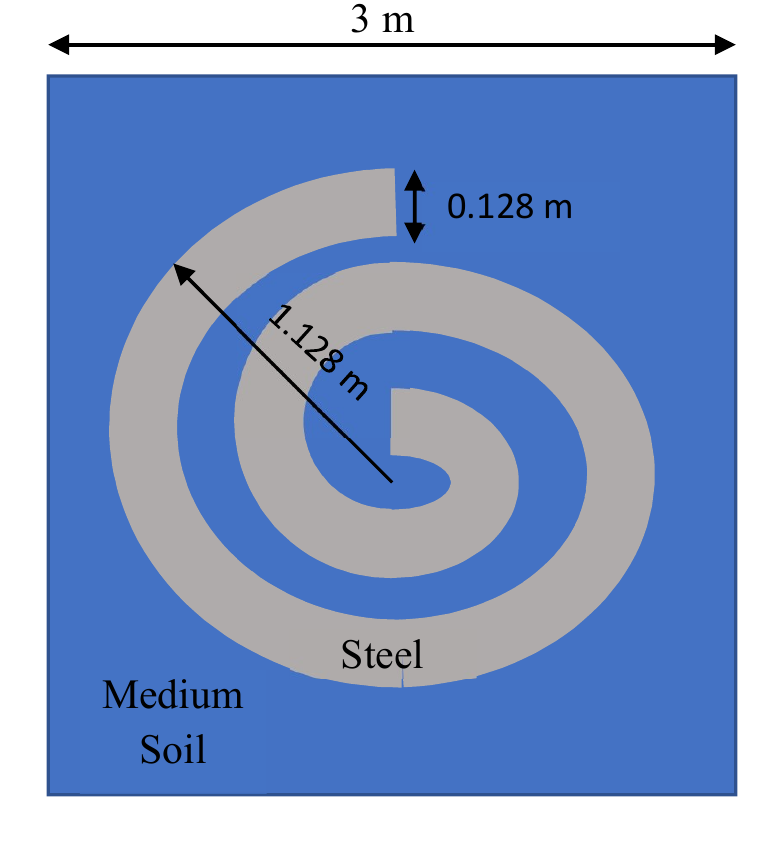}
		\caption{Cross-section of cylindrical swiss roll-like inclusion with thickness 0.128 m} 
		\label{F211a} 
		\vspace{2ex}
	\end{subfigure}
	\begin{subfigure}[b]{0.7\linewidth}
		\centering
		\includegraphics[width=1.0\linewidth]{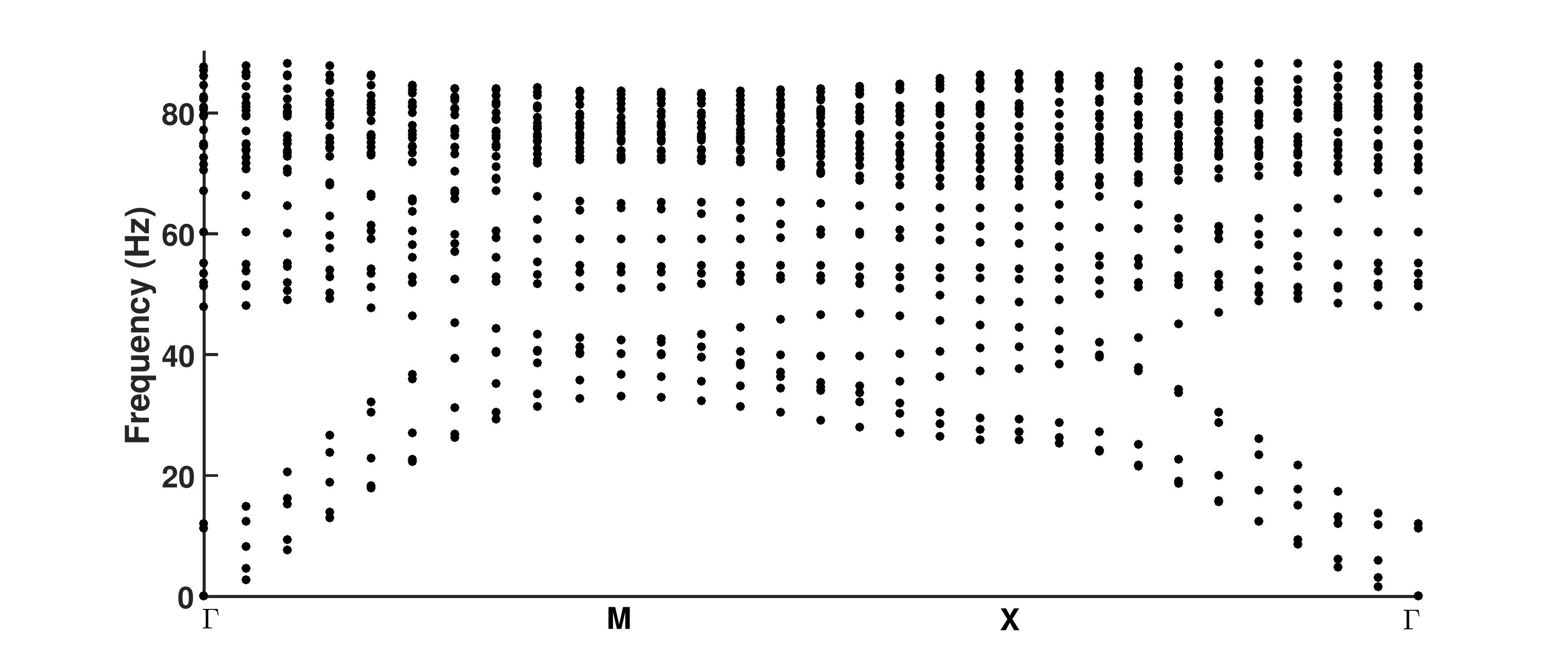} 
		\caption{Dispersion curves for cylindrical swiss roll inclusion with 0.128 m thickness} 
		\label{F211b} 
		\vspace{2ex}
	\end{subfigure} 
	\begin{subfigure}[b]{0.3\linewidth}
		\centering
		\includegraphics[width=0.8\linewidth]{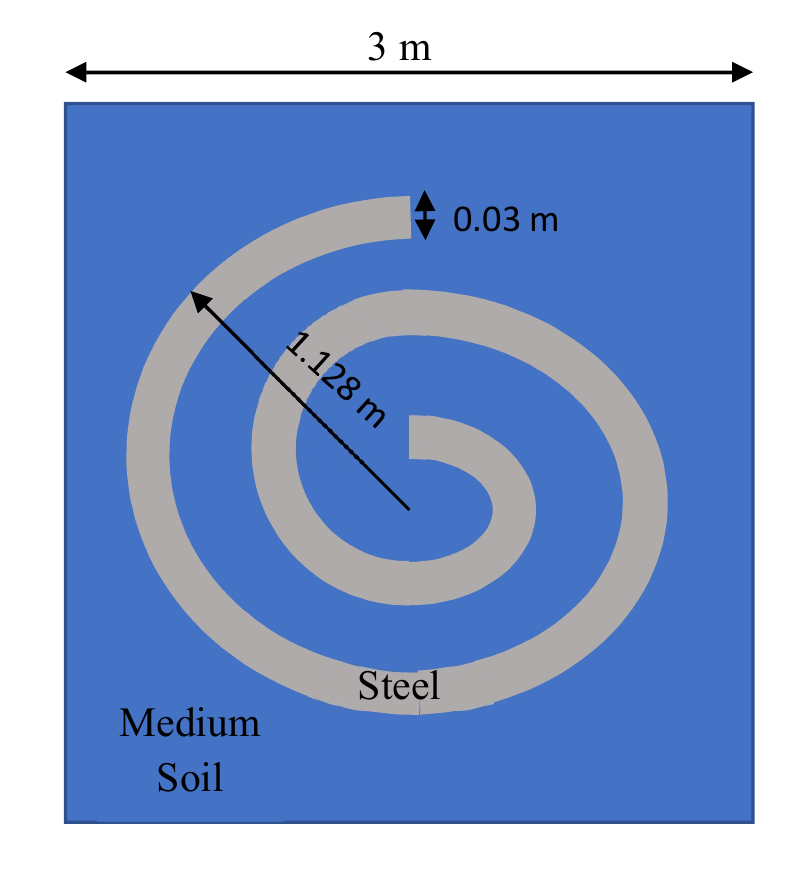} 
		\caption{Cross-section of cylindrical swiss roll inclusion with thickness 0.03 m} 
		\label{F211c} 
		\vspace{-0ex}
	\end{subfigure}
	\begin{subfigure}[b]{0.7\linewidth}
		\centering
		\includegraphics[width=1\linewidth]{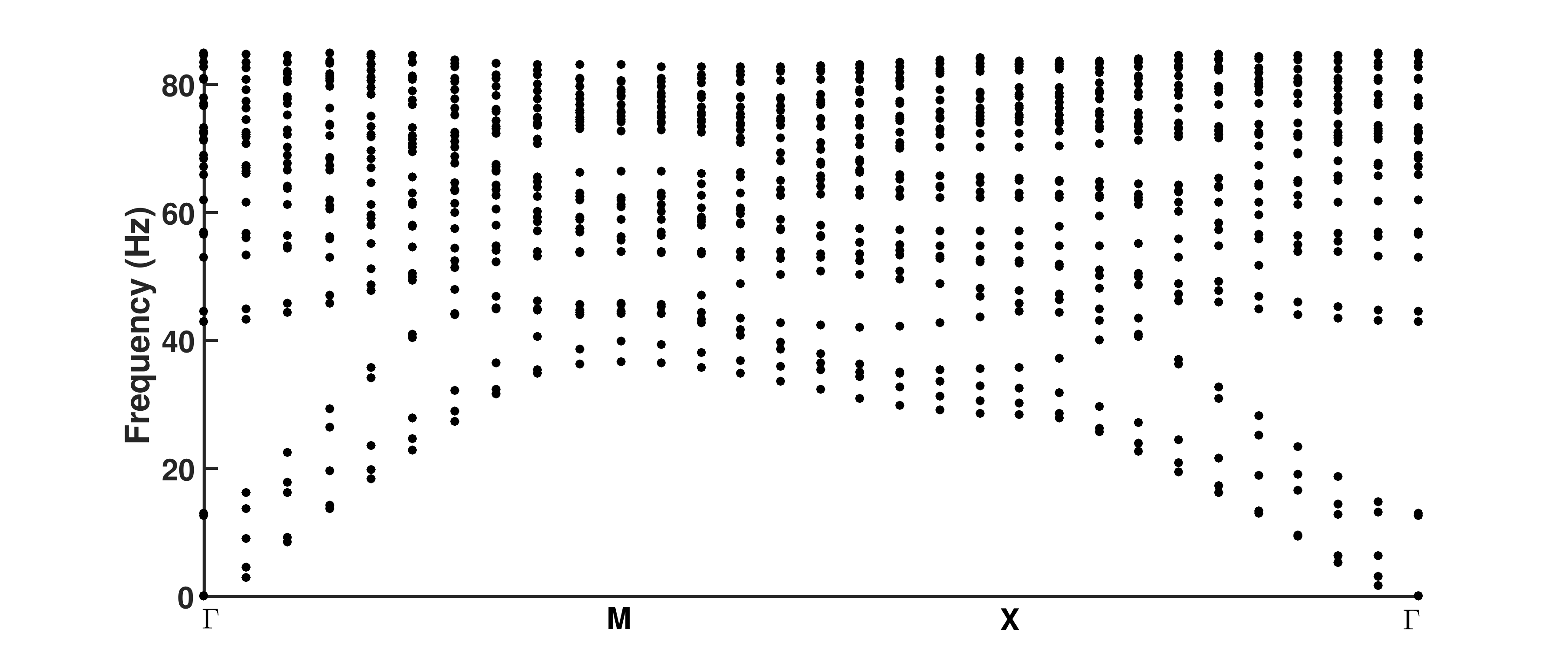} 
		\caption{Dispersion curves for cylindrical swiss roll-like inclusion with 0.03 m thickness} 
		\label{F211d} 
	\end{subfigure}
	\caption{Unclamped configuration of cylindrical swiss roll like geometry with different thickness of steel inclusions in the elementary cell (3m x3m x10m) of periodic media (medium soil)  . Diameter of swiss roll is considered same as solid circular inclusions in Fig.\ref{F1}. No bandgap is observed in both the cases}
	\label{F211} 
\end{figure}

\begin{figure}[H] 
	\begin{subfigure}[b]{0.55\linewidth}
		\centering
		\includegraphics[width=1.0\linewidth]{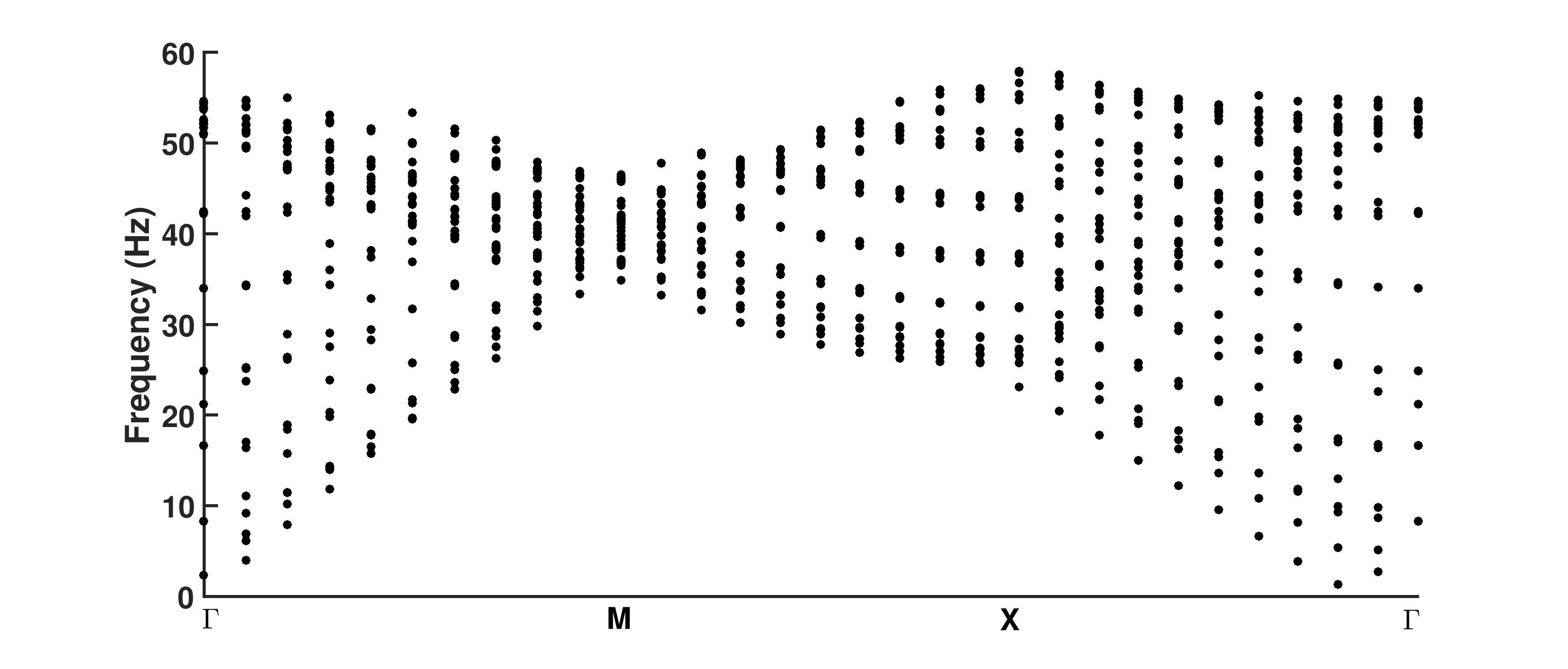}
		\caption{Dispersion curve for Labyrithine microstructure with rubber as inclusion} 
		\label{F311a} 
		\vspace{2ex}
	\end{subfigure}
	\begin{subfigure}[b]{0.55\linewidth}
		\centering
		\includegraphics[width=1.0\linewidth]{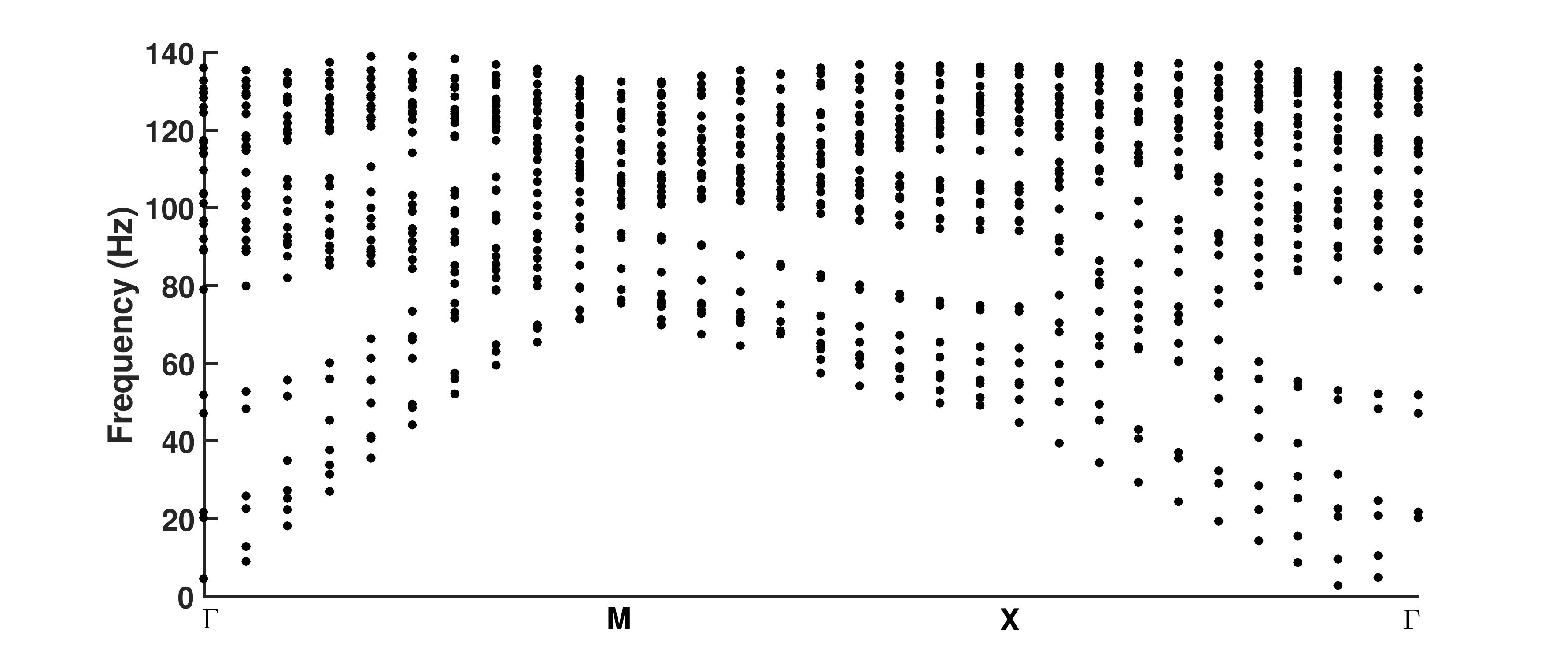} 
		\caption{Dispersion curve for Labyrithine microstructure with concrete as inclusion} 
		\label{F311b} 
		\vspace{2ex}
	\end{subfigure} 
	\caption{Dispersion curves for unclamped configuration of Labyrithine geometry as shown in fig. \ref{F11a} with different constituent material, i.e., (a) rubber and (b) concrete. No bandgap is observed with these inclusions}
	\label{F311} 
\end{figure}

Motivated by the electromagnetic metamaterials literature, 
 other investigated microstructure geometries are now considered:  coil/Labyrinthine (complex sheet piling type of inclusion), split ring-like and swiss roll-like cylindrical steel inclusions in medium soil (see Fig. \ref{F111} and \ref{F211}).  The same substitution ratio, i.e., 0.445 was considered as in the former two microstructures, whereas the swiss roll-like inclusion is considered with a diameter equal to that of the solid circular inclusion in Fig.\ref{F1}, i.e., 1.128 m with a substitution ratio of 0.15. Notably, in medium soil, without clamping the bottom, a split-ring like inclusion with 4 gaps shows a bandgap of approximately 10.6 Hz, whereas the Labyrinthine inclusion shows no bandgap. Testing with other material constituents for a Labyrithine inclusion (Fig. \ref{F11a}), i.e., with rubber and concrete  no bandgap was observed (see Fig. \ref{F311}). On the other hand,  split-ring with 2 gaps shows an enhanced bandgap (12 Hz in the range of 46.84 Hz- 59.23 Hz) as compared to 4 gap split-ring (10.6 Hz in the range of 48.53 Hz- 58.35 Hz) with same substitution ratio (0.445). Dispersion curves for swiss roll-like steel inclusion with different thickness, i.e., 0.128 m and 0.03 m in medium soil (3 m $\times$ 3 m $\times$ 10 m) is obtained by without clamping the bottom and is shown in Fig.\ref{F211}. Clearly, there is no bandgap for these configurations. 

\begin{figure}[H] 
	\begin{subfigure}[b]{1\linewidth}
		\centering
		\includegraphics[width=0.7\linewidth]{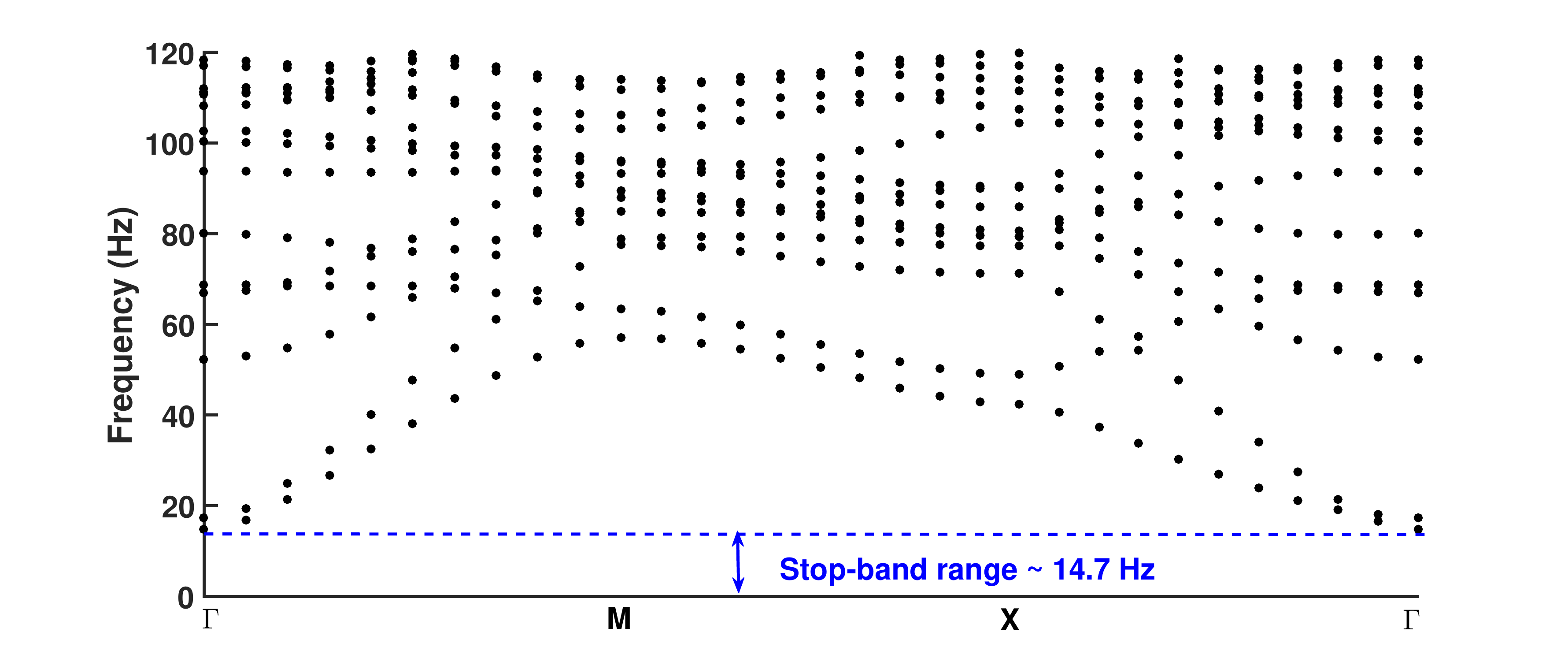} 
		\caption{Dispersion curves for cylindrical Labyrithine inclusion} 
		\label{F112a} 
	\end{subfigure} 
	\begin{subfigure}[b]{1\linewidth}
		\centering
		\includegraphics[width=0.7\linewidth]{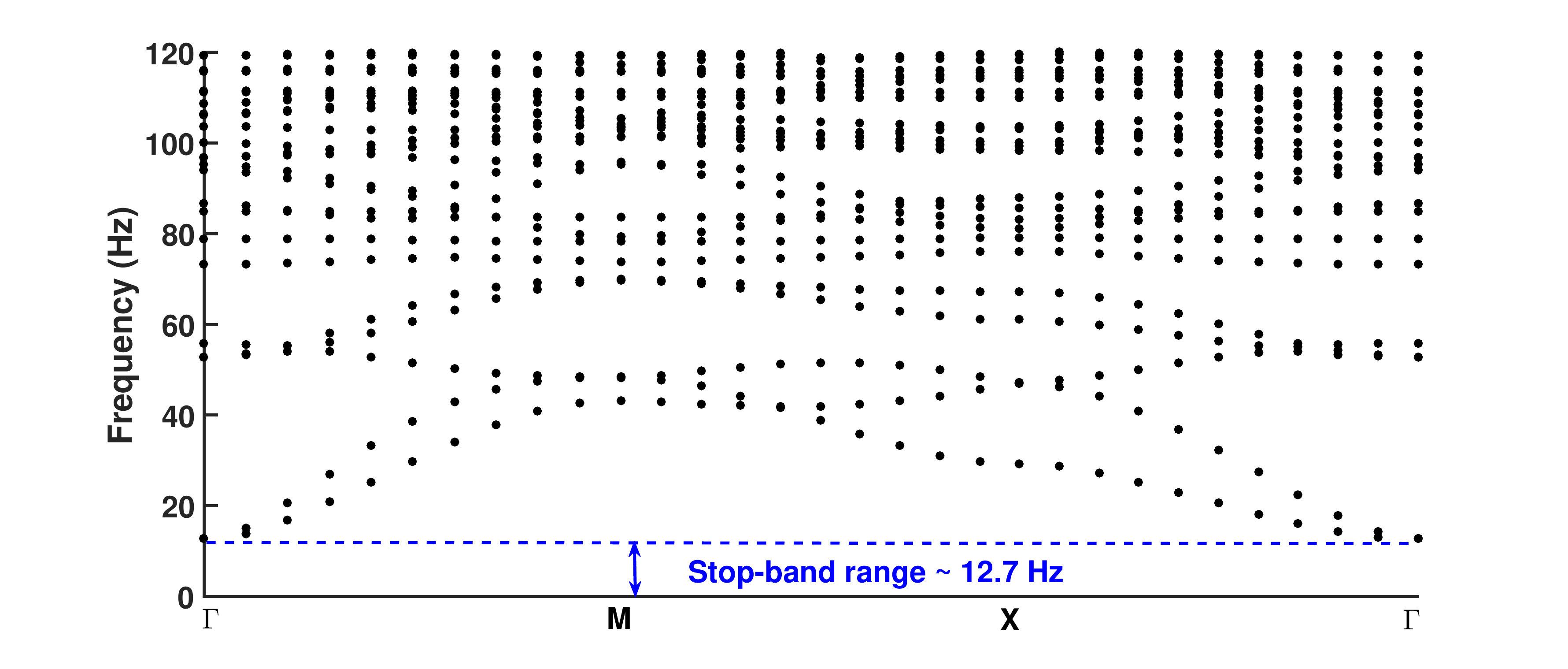} 
		\caption{Dispersion curves for cylindrical split-ring-like inclusion} 
		\label{F112b} 
	\end{subfigure} 
	\begin{subfigure}[b]{1\linewidth}
		\centering
		\includegraphics[width=0.7\linewidth]{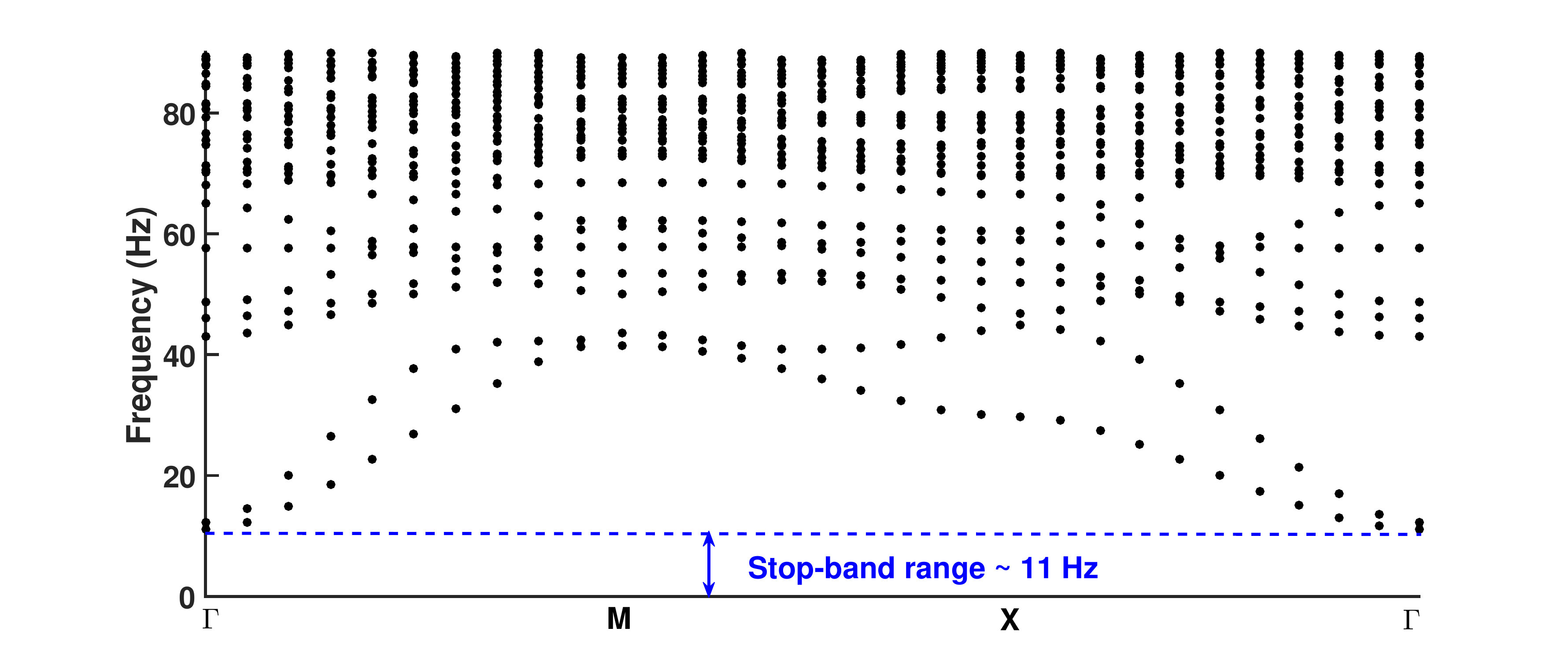} 
		\caption{Dispersion curves for cylindrical swiss-roll-like inclusion} 
		\label{F112c} 
	\end{subfigure} 
	\caption{Dispersion curves for clamped configuration with steel as inclusion material for (a) Labyrithine, (b) 4 gaps split ring, and (c) swiss roll  with 0.128 m thickness with  substitution ratio as 0.445 (for swiss roll: 0.15).}
	\label{F112} 
\end{figure}

The stop-bands for the configurations in Fig.\ref{F111} and \ref{F211} are obtained by clamping the bottom and are shown in Fig.\ref{F112}. It is observed that the Labyrinthine inclusion gives higher stop-bands in contrast to other two, i.e., 14.7 Hz for the Labyrinthine steel inclusion in comparison with 12.7 Hz and 11 Hz for split-ring and swiss-roll like steel inclusions. 

\begin{figure}[H] 
	\begin{subfigure}[b]{1\linewidth}
		\centering
		\includegraphics[width=0.7\linewidth]{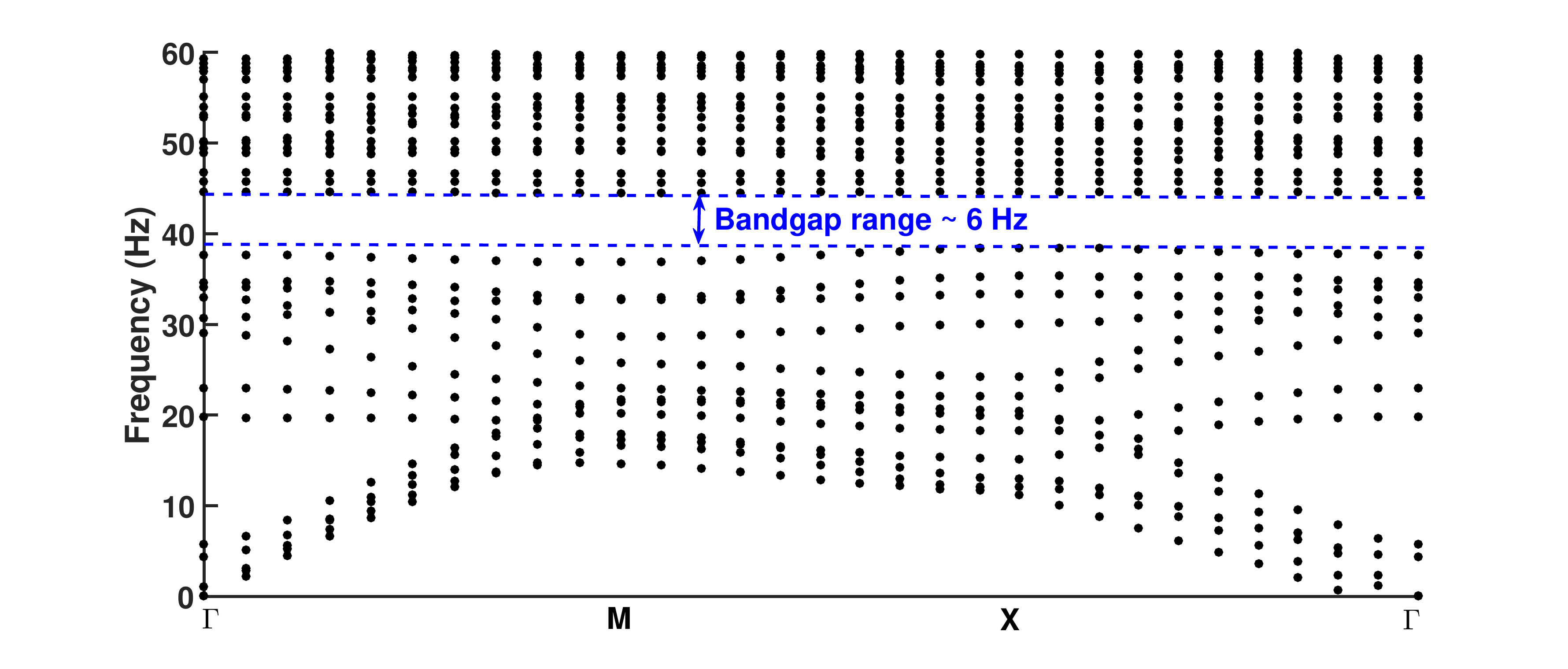} 
		\caption{Dispersion curves for cylindrical Labyrithine inclusion in very soft clay type 1 (cross-section details in Fig. \ref{F11a})} 
		\label{F113a} 
	\end{subfigure} 
	\begin{subfigure}[b]{1\linewidth}
		\centering
		\includegraphics[width=0.7\linewidth]{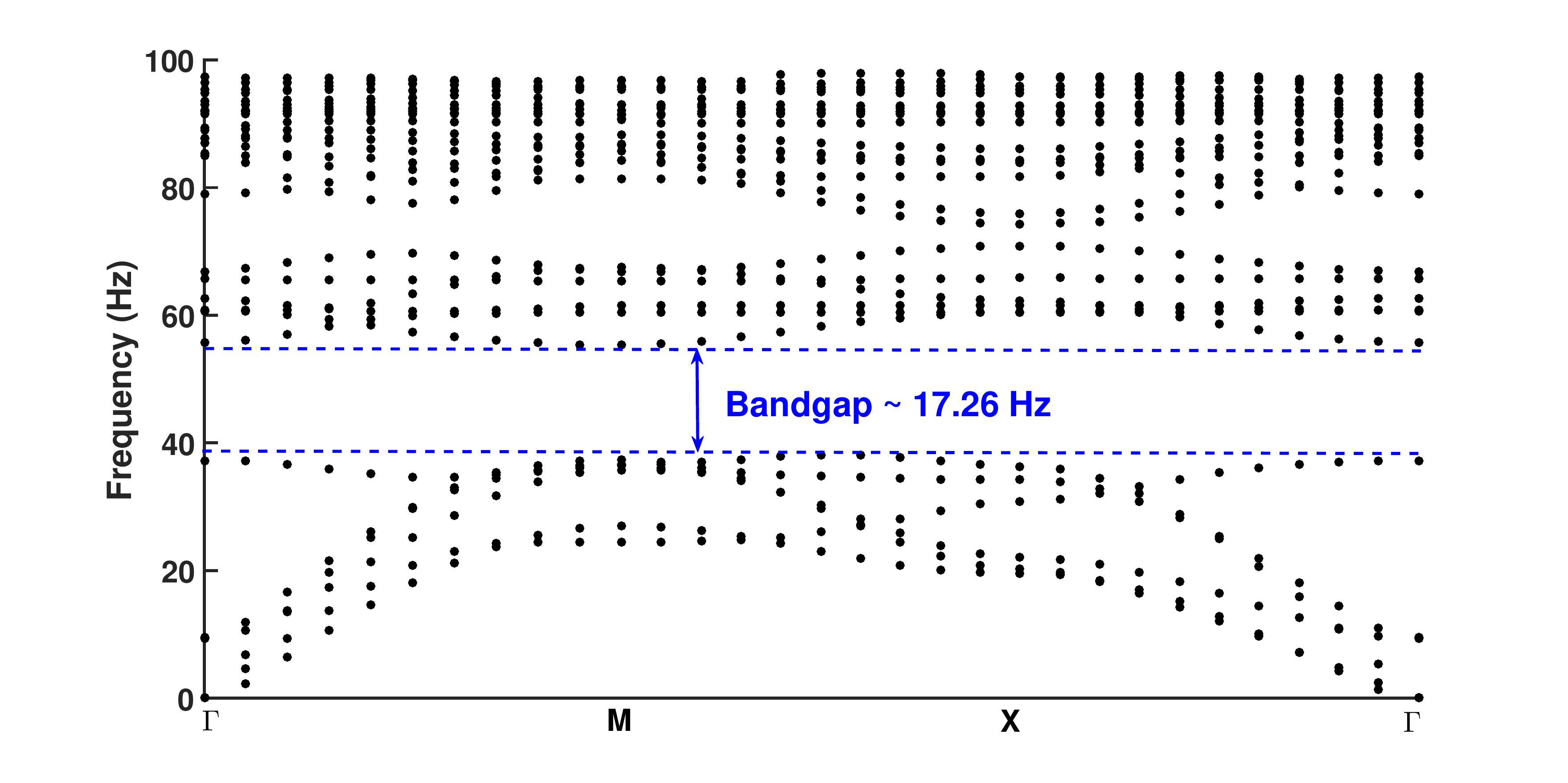} 
		\caption{Dispersion curves for 2 gap cylindrical split-ring-like inclusion in soft soil (cross-section details in Fig. \ref{F11e})} 
		\label{F113b} 
	\end{subfigure} 
	\begin{subfigure}[b]{1\linewidth}
		\centering
		\includegraphics[width=0.7\linewidth]{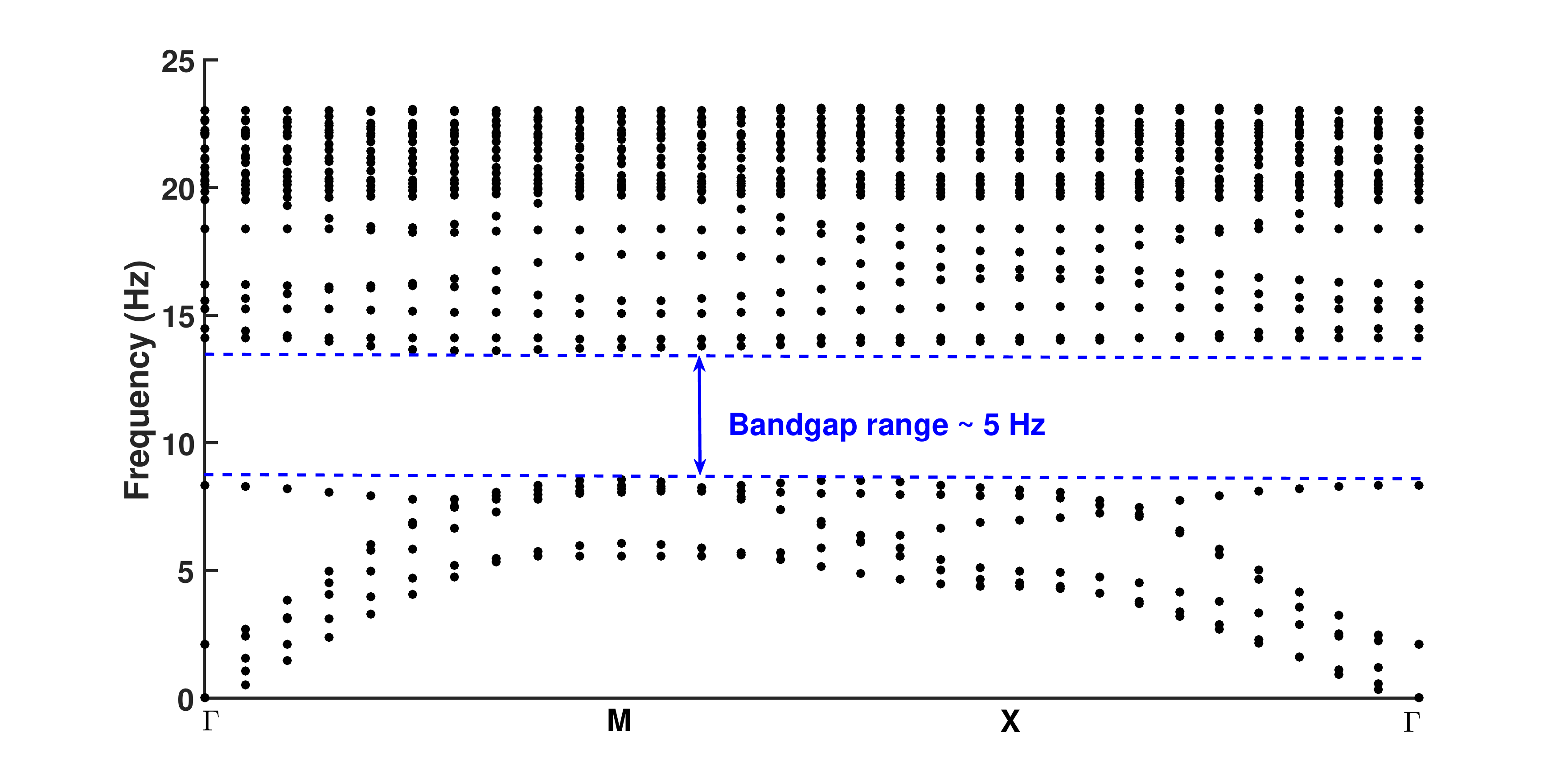} 
		\caption{Dispersion curves for 2 gap cylindrical split-ring-like inclusion in very soft clay type 2 (cross-section details in Fig. \ref{F11e})} 
		\label{F113c} 
	\end{subfigure} 
	\caption{Dispersion curves for unclamped configurations of Labyrithine and 2 gap split ring steel inclusions with 0.445 sustitution ratio in soft ($E=96.5 \ \text{MPa}$, $\mu=0.33$ and $\rho=1650 \  \text{kg}/\text{m}^3$), very soft clay type 1 ($E=10 \ \text{MPa}$, $\mu=0.25$ and $\rho=1400 \  \text{kg}/\text{m}^3$) and very soft clay type 2 ($E=5 \ \text{MPa}$, $\mu=0.35$ and $\rho=1633 \  \text{kg}/\text{m}^3$) .}
	\label{F113} 
\end{figure} 

\begin{figure}[H] 
	\begin{subfigure}[b]{0.3\linewidth}
		\centering
		\includegraphics[width=1\linewidth]{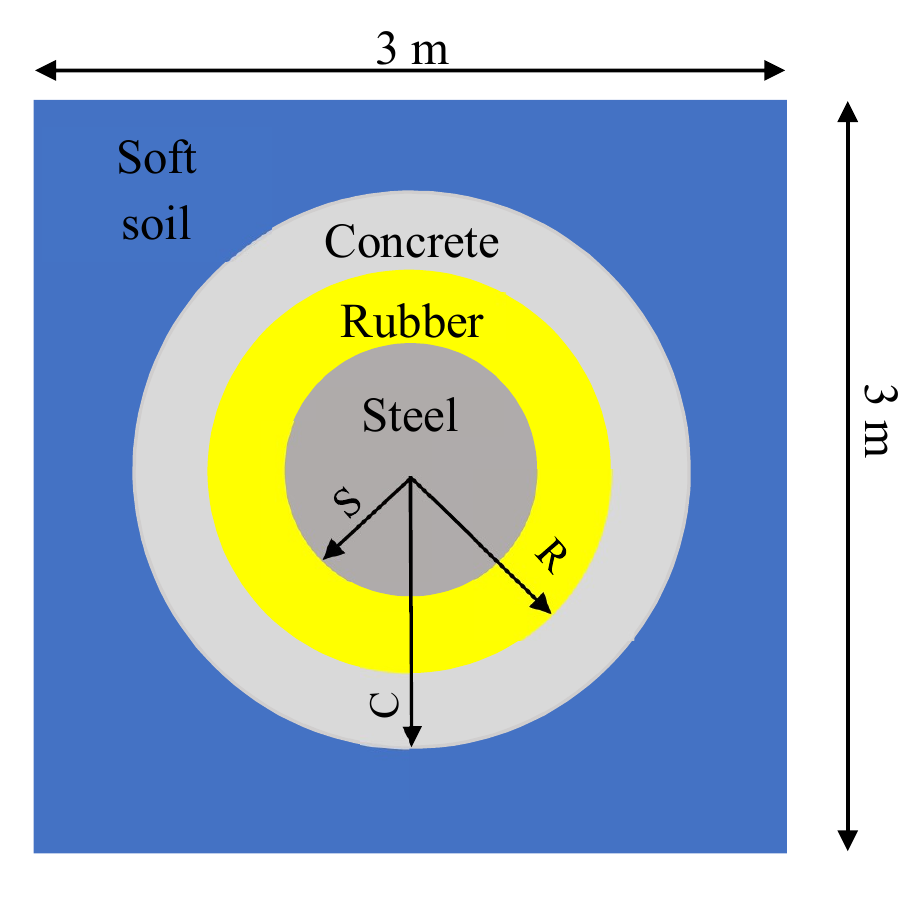} 
		\caption{Cross-section of triphasic inclusion in soft soil} 
		\label{F114a} 
	\end{subfigure} 
	\begin{subfigure}[b]{0.7\linewidth}
		\centering
		\includegraphics[width=1\linewidth]{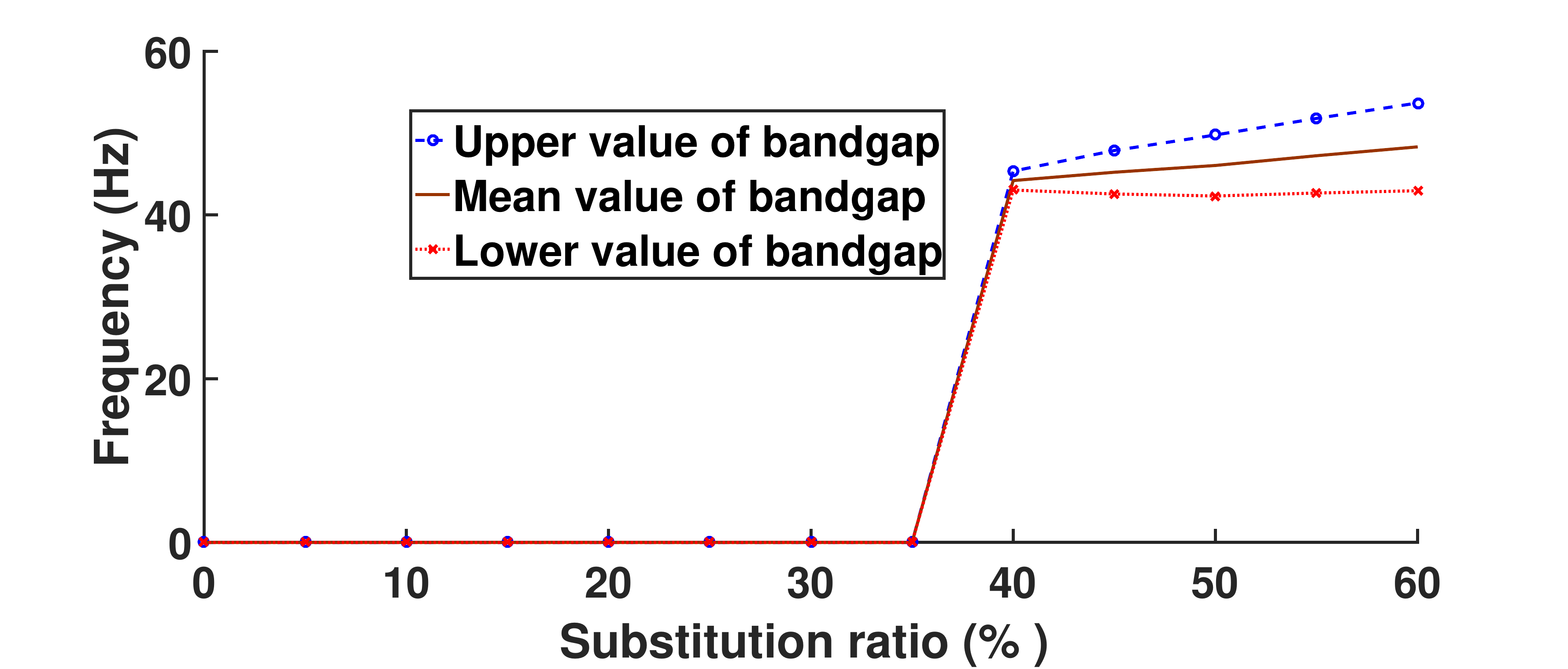} 
		\caption{Variation of bandgap with substitution ratio of inclusions} 
		\label{F114b} 
	\end{subfigure} 
	\caption{Unclamped configuration (all three type of inclusion, i.e., concrete, rubber and steel), and Variation of bandgap width versus substitution ratio of triphasic inclusion in soft soil. The ratio of radius of each inclusion material for different substitution ratio is kept same, i.e., $\frac{S}{R} \ = \ \frac{R}{C} \ = \ 0.83$, where S, R and C denotes radius of steel, rubber and concrete, respectively. The dimension of unit cell is 3 m $\times$ 3 m $\times$ 10 m and properties of soft soil is taken as $E=96.5 \ \text{MPa}$, $\mu=0.33$ and $\rho=1650 \  \text{kg}/\text{m}^3$.}
	\label{F114} 
\end{figure} 

We now investigating the effect of different soil properties, i.e., with soft soil having properties $E=96.5 \ \text{MPa}$, $\mu=0.33$ and $\rho=1650 \  \text{kg}/\text{m}^3$  and very soft soil type 1 with properties $E=10 \ \text{MPa}$, $\mu=0.25$ and $\rho=1400 \  \text{kg}/\text{m}^3$. The  Labyrithine and 2 gap split ring-like steel inclusions show an improvement on the width of the bandgaps obtained via dispersion curves in Fig. \ref{F111}. The Labyrithine steel inclusion shows a bandgap of 6 Hz in the range of 38 Hz - 44 Hz in  soft soil (Fig. \ref{F113a}) in contrast to medium soil (Fig.\ref{F11b}), which shows no bandgap. Split ring with 2 gaps shows an enhanced bandgap (approx. 17 Hz) in soft soil with soil properties $E=96.5 \ \text{MPa}$, $\mu=0.33$ and $\rho=1650 \  \text{kg}/\text{m}^3$ (Fig. \ref{F113b}). The bandgap is also obtained in a lower region than that obtained in the medium soil (Fig.\ref{F11f}). On further exploring the soil properties, the 2 gap split ring shows an excellent improvement over the bandgap in a very soft soil type 2 with soil properties $E=5 \ \text{MPa}$, $\mu=0.35$ and $\rho=1633 \  \text{kg}/\text{m}^3$. A bandgap of 5 Hz is obtained at a much lower region (approx. 8 Hz - 13 Hz) in very soft soil type 2 for 2 gap split ring (Fig. \ref{F113c}). Finally, we investigate a typical configuration such as a triphasic inclusion as shown in Fig. \ref{F114} \cite{duplan2014prediction}. A circular cylinder of steel, surrounded by a thick rubber shell, surrounded by a concrete outer shell within a bulk of soft soil with parameters $E=96.5 \ \text{MPa}$, $\mu=0.33$ and $\rho=1650 \  \text{kg}/\text{m}^3$ as shown in Fig. \ref{F114a}.  The ratio of radius of each inclusion material is taken such as ${S}/{R} \ = \ {R}/{C} \ = \ 0.83$, where $S, R$ and $C$ denote radius of steel, rubber and concrete, respectively. The dimension of the unit cell is 3 m $\times$ 3 m $\times$ 10 m. With increasing substitution ratio of material constituents, the varying bandgap width is plotted as shown in Fig. \ref{F114b}. It can be seen that the band gap for triphasic inclusion appears at lower substitution rates than for monophasic substitution cases.              

\subsection{Effect of varying orientation in square inclusions}
\label{S32}
\begin{figure}[H] 
	\begin{subfigure}[b]{0.5\linewidth}
		\centering
		\includegraphics[width=0.6\linewidth]{Square_CS.pdf} 
		\caption{Cross-section of $0\degree$ orientation of square $\\$ inclusion with 0.445 substitution ratio} 
		\label{F2a} 
	\end{subfigure}
	\begin{subfigure}[b]{0.5\linewidth}
		\centering
		\includegraphics[width=0.6\linewidth]{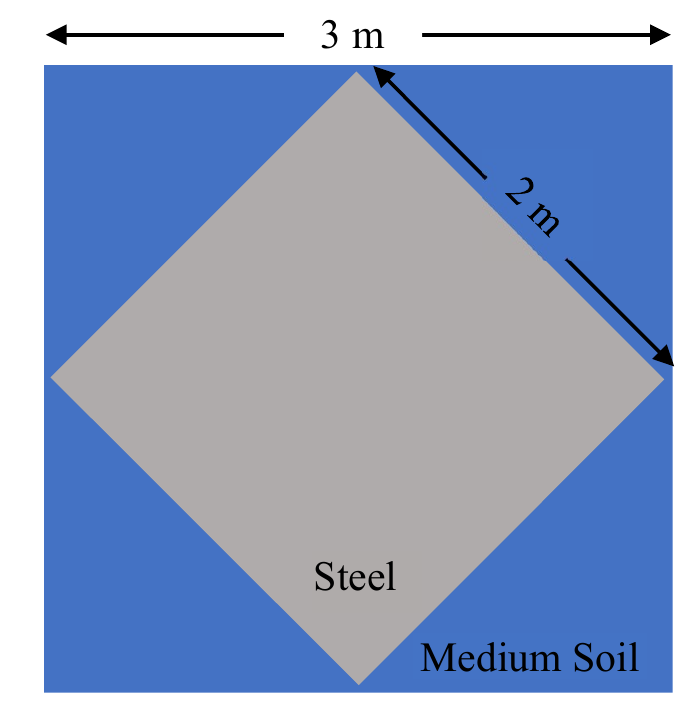} 
		\caption{Cross-section of $45\degree$ orientation of square  $\\$ inclusion with 0.445 substitution ratio} 
		\label{F2b} 
	\end{subfigure} 
	\begin{subfigure}[b]{1\linewidth}
		\centering
		\includegraphics[width=1\linewidth]{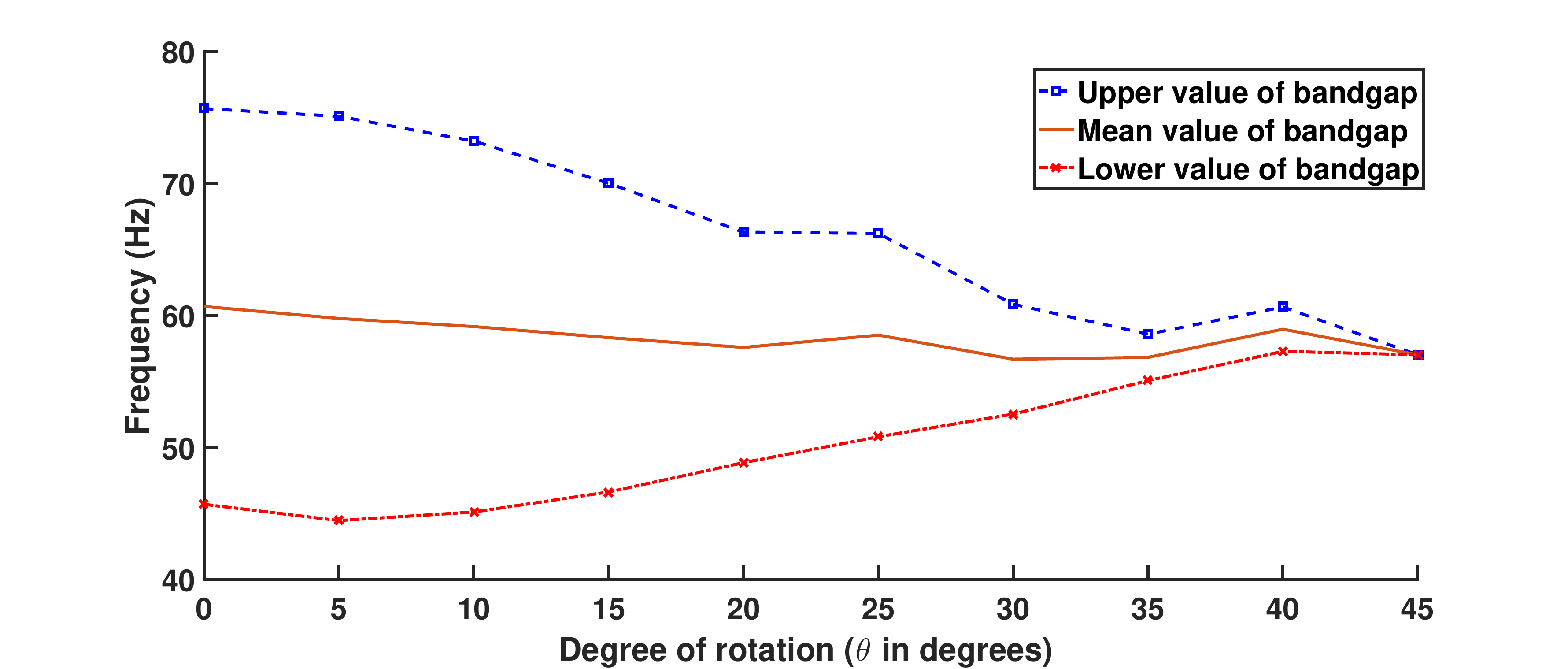} 
		\caption{Bandgap for unclamped configuration versus orientation angle (degree) plot for square inclusion (steel) in medium soil at same substitution ratio, i.e., 0.445} 
		\label{F2c} 
	\end{subfigure} 
	\caption{ Effect of change in orientation angle (in degrees) of unclamped regular-shaped square steel inclusion (2 m $\times$ 2 m $\times$ 10 m) with 0.445 susbtitution ratio in medium soil (3 m $\times$ 3 m $\times$ 10 m ; $E=153 \ \text{MPa}$, $\mu=0.33$ and $\rho=1800 \  \text{kg}/\text{m}^3$)  on bandgap; (a) and (b) shows the cross-section of $0\degree$ and $45\degree$ angle orientation of unclamped square inclusion; (c) shows the declining trend in bandgap with increase in angle from $0\degree$ to $45\degree$ in anticlockwise direction}
	\label{F2} 
\end{figure}

\begin{figure}[H]
	\begin{center}
		{\includegraphics[width=1\textwidth]{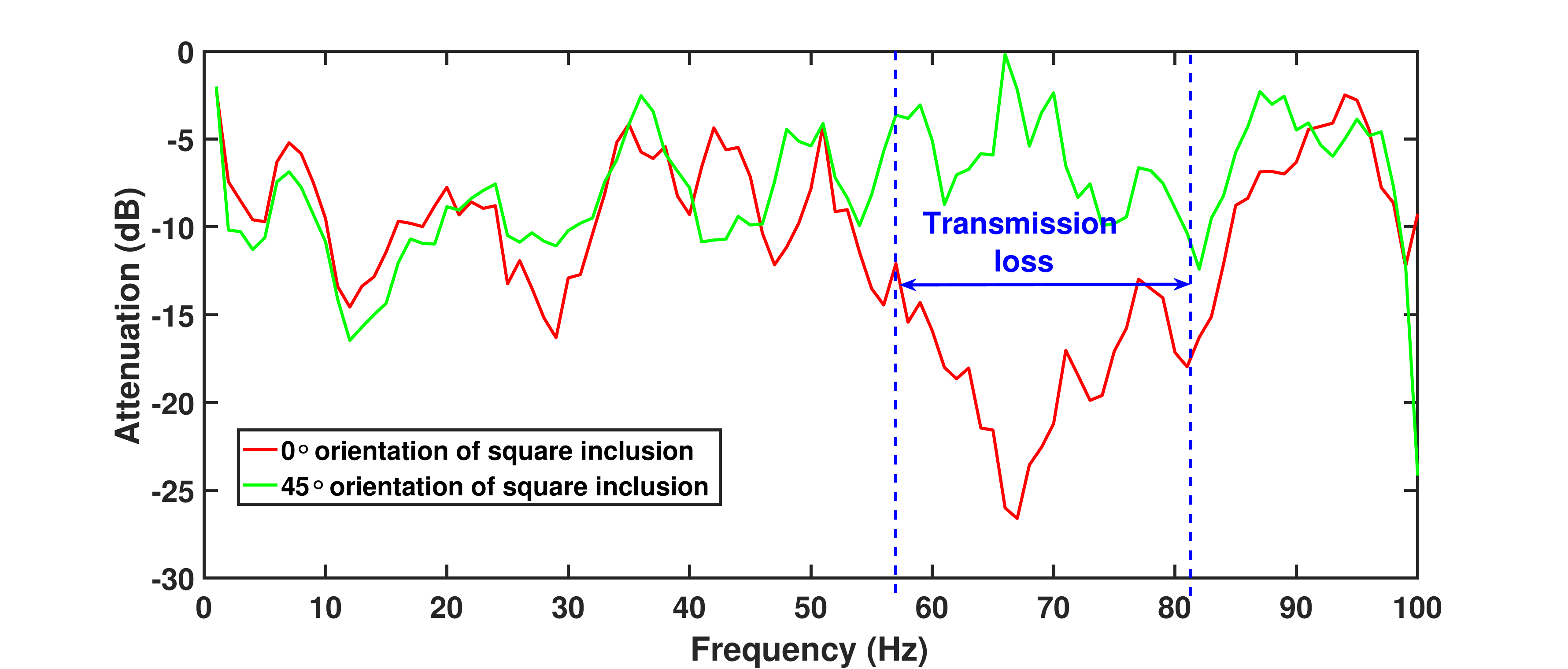}}
		{\caption{\label{F3} Transmission spectra of $0\degree$ and $45\degree$ orientation of square steel inclusion in medium soil matrix  (bottom remains unclamped) is computed using a finite array of metamaterial design (Corresponding to Fig. \ref{F2}). A dip is observed at approximately 68 Hz, representing large wave attenuation.}}
	\end{center}
\end{figure}

Having verified in subsection \ref{S31}, that  using a square geometry as the inclusion leads to better performance, here we explore the effect of varying its orientation within the microstructure of periodic media by rotating it from $0\degree$ to $45\degree$. With this exercise, we further explore the microstructure geometries towards achieving higher bandgaps. Steel inclusions are considered as the column with regular-shaped square geometry ($2 \ \text{m}\times2 \ \text{m}\times10 \ \text{m}$) within an unstructured medium soil ($3 \ \text{m}\times3 \ \text{m}\times10 \ \text{m}$), with properties same as in section \ref{S31}. The orientation of the square inclusion is made to vary from $0\degree$ to $45\degree$, in increments of $5\degree$.  Fig. \ref{F2c}  shows the plot for bandgaps obtained from varying the orientation of inclusion plotted against the angle. Interestingly, the plot shows a declining trend for the bandgap with increase in angle, starting from 28 Hz at $0\degree$ to 0 Hz at $45\degree$. Thus, in summary a square inclusion oriented at $0\degree$ gives a wider bandgap in comparison to other orientations when examined with same substitution ratio. To further investigate the efficacy of the design, transmission losses (see Fig. \ref{F3}) are computed over a finite array of SM design consisting of five square steel columns (oriented at $0\degree$ and $45\degree$, respectively). For the computation, Rayleigh wave is incident normally as a line source. With $0\degree$ orientation a dip is observed in the range of 58 Hz-82 Hz, which shows possible large wave attenuation/transmission loss, whereas no such observations is made for the $45\degree$ case.

\subsection{Feasibility of inclusion material}
\label{S33}
\begin{figure}[H] 
	\begin{subfigure}[b]{0.3\linewidth}
		\centering
		\includegraphics[width=0.8\linewidth]{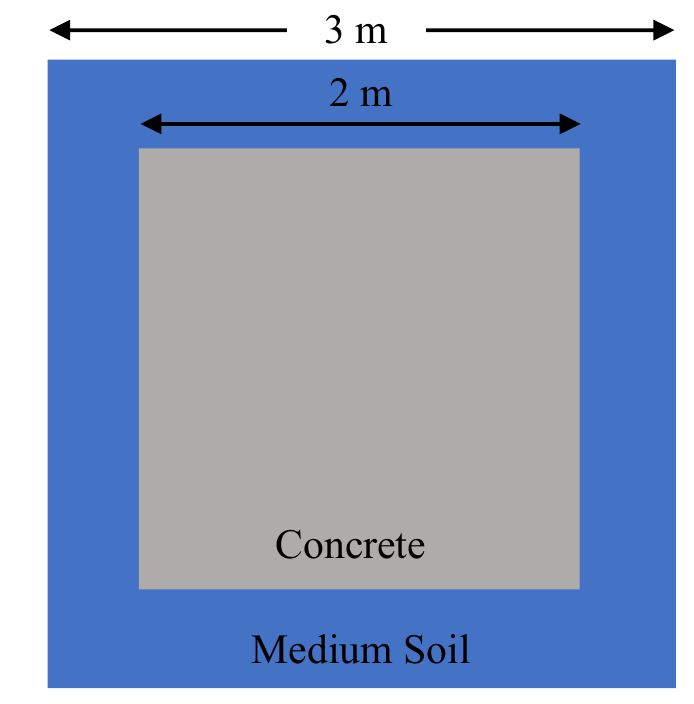} 
		\caption{Concrete inclusion in medium soil matrix with 0.445 susbtitution ratio} 
		\label{F4a} 
		\vspace{2ex}
	\end{subfigure}
	\begin{subfigure}[b]{0.7\linewidth}
		\centering
		\includegraphics[width=1.0\linewidth]{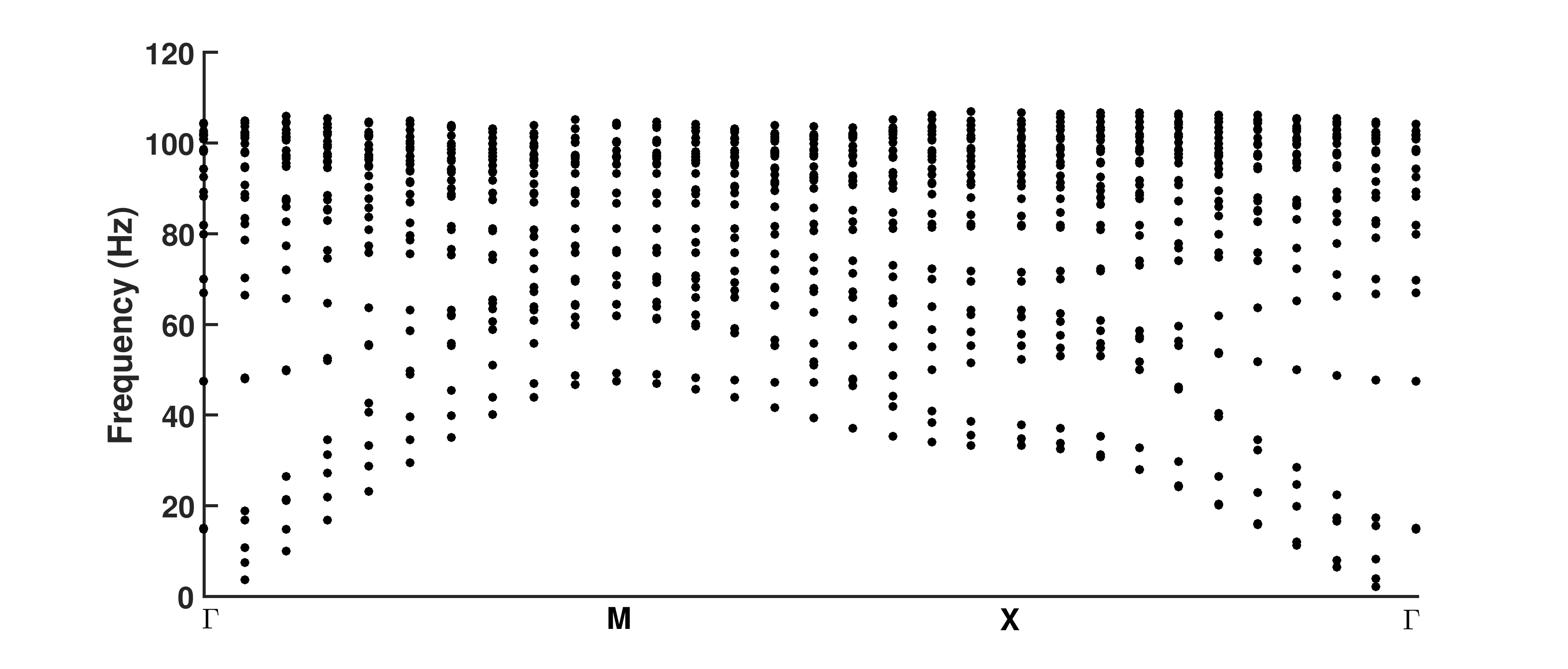} 
		\caption{Dispersion curves for square inclusion in medium soil} 
		\label{F4b} 
		\vspace{4ex}
	\end{subfigure} 
	\begin{subfigure}[b]{0.3\linewidth}
		\centering
		\includegraphics[width=0.8\linewidth]{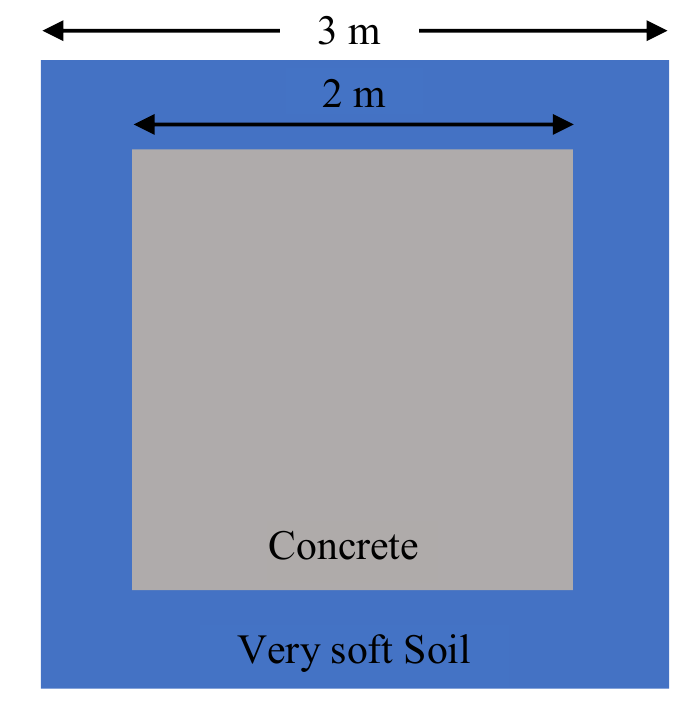} 
		\caption{Concrete inclusion in very soft soil matrix with 0.445 susbtitution ratio} 
		\label{F4c} 
		\vspace{-1ex}
	\end{subfigure}
	\begin{subfigure}[b]{0.7\linewidth}
		\centering
		\includegraphics[width=1\linewidth]{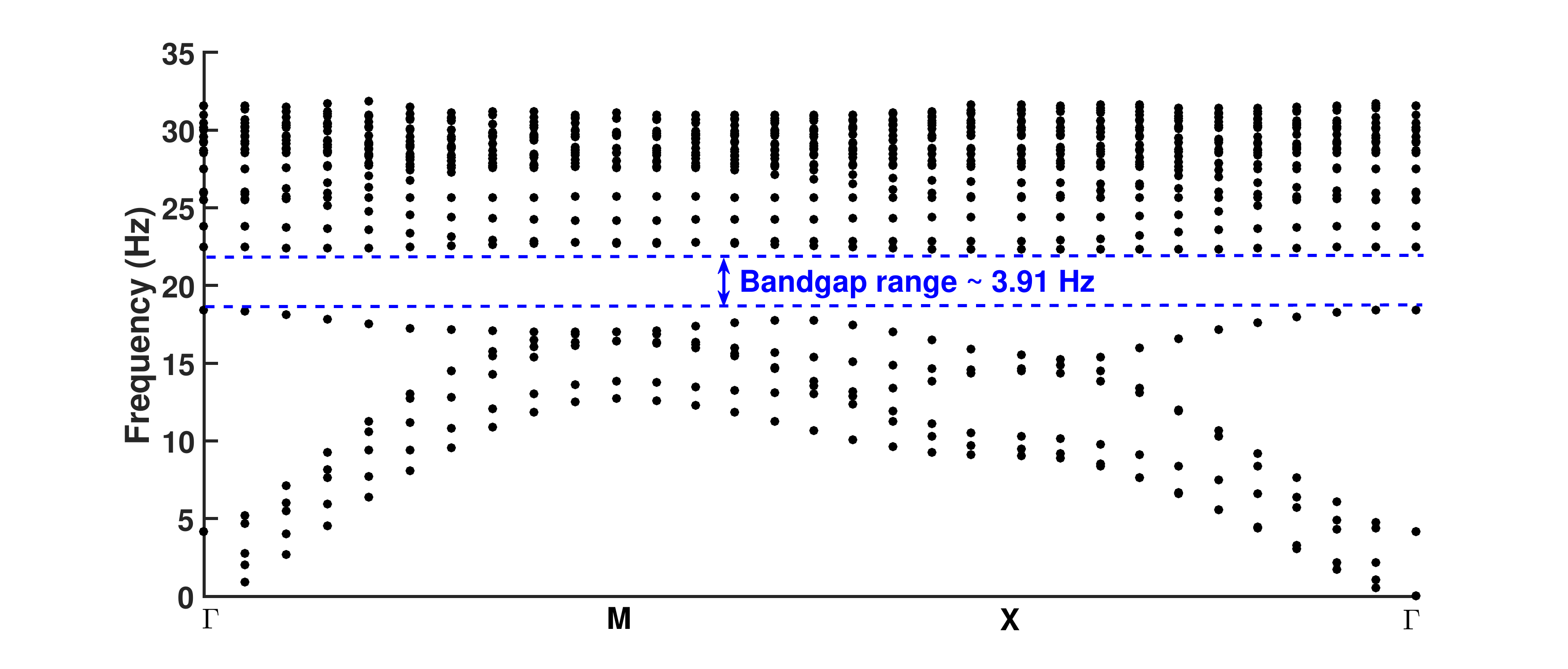} 
		\caption{Dispersion curves for square inclusion in loose soil} 
		\label{F4d} 
	\end{subfigure} 
	\caption{Unclamped configuration of concrete inclusion (with 0.445 substitution ratio) in (a)  medium soil ($E=153 \ \text{MPa}$, $\mu=0.33$ and $\rho=1800 \  \text{kg}/\text{m}^3$); (c) very soft soil ($E=10 \ \text{MPa}$, $\mu=0.25$ and $\rho=1400 \  \text{kg}/\text{m}^3$ ). Corresponding dispersion curves obtained are shown in (b) and (d), around the edges of the irreducible Brillouin zone $\Gamma X M$.  No bandgap is observed with medium soil matrix; whereas in very soft soil a bandgap, approximately 3.91 Hz is observed.}
	\label{F4} 
\end{figure}

\begin{figure}[H]
	\begin{center}
		{\includegraphics[width=1\textwidth]{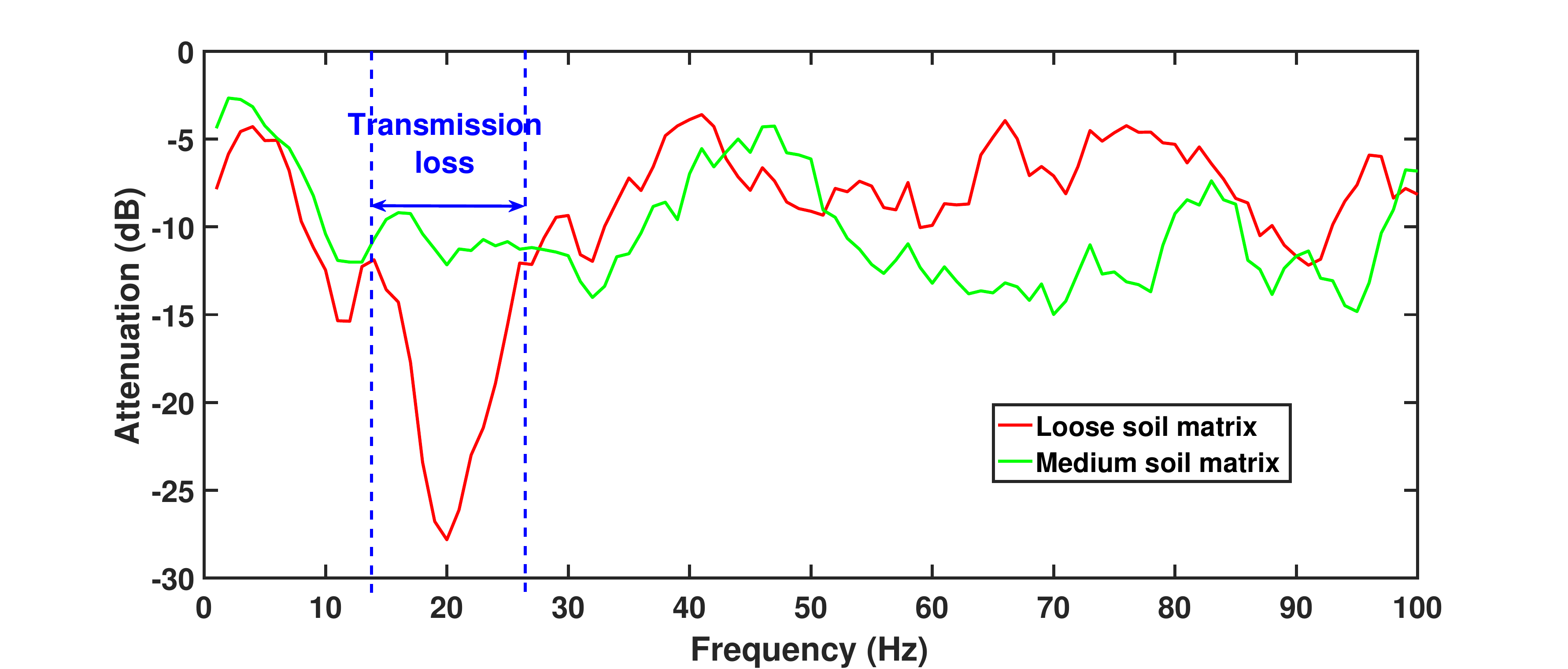}}
		{\caption{\label{F5}  Transmission losses for unclamped concrete inclusions (2 m $\times$ 2 m $\times$ 10 m) in very soft soil type 1 ($\text{E}=10 \ \text{MPa}$, $\mu=0.25$ and $\rho=1400 \  \text{kg}/\text{m}^3$) and medium soil  ($\text{E}=153 \ \text{MPa}$, $\mu=0.33$ and $\rho=1800 \  \text{kg}/\text{m}^3$) matrix (3 m $\times$ 3 m $\times$ 10 m) corresponding to Fig. \ref{F4}.  The substitution ratio is kept same, i.e., 0.445. Notably, square concrete inclusions  in very soft soil type 1 shows attenuation in the bandgap range, i.e., 18.45 Hz-22.36 Hz.}}
	\end{center}
\end{figure}

\begin{figure}[H]
	\begin{center}
		{\includegraphics[width=1\textwidth]{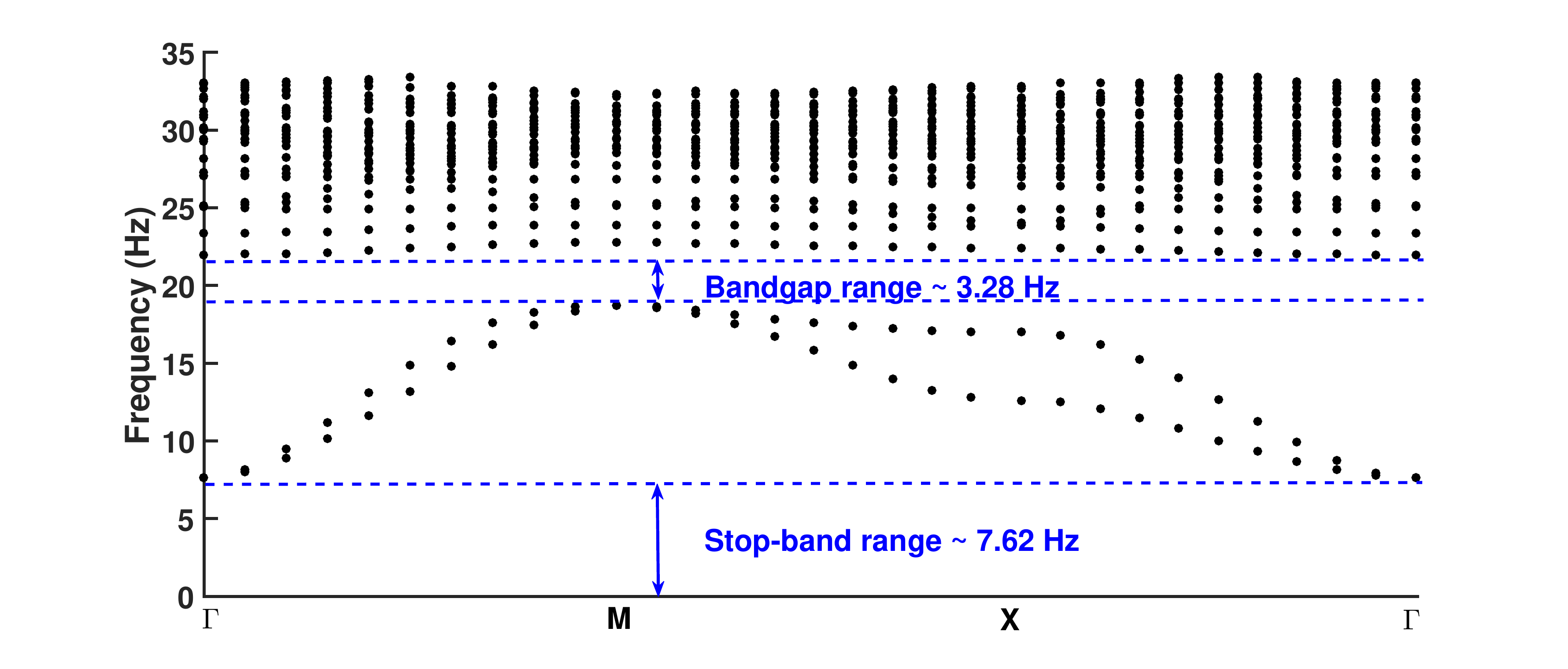}}
		{\caption{\label{F6} Dispersion curve for clamped configuration of M50 grade square concrete columns (2 m $\times$ 2 m $\times$ 10 m) in soft soil matrix (3 m $\times$ 3 m $\times$ 10 m; $E=10 \ \text{MPa}$, $\mu=0.25$ and $\rho=1400 \  \text{kg}/\text{m}^3$) (substitution ratio 0.445). Two bandgaps were observed; a stopband of 7.62 Hz and a bandgap in the range of 18.72-22 Hz}}
	\end{center}
\end{figure}

\begin{figure}[H]
	\begin{center}
		{\includegraphics[width=1\textwidth]{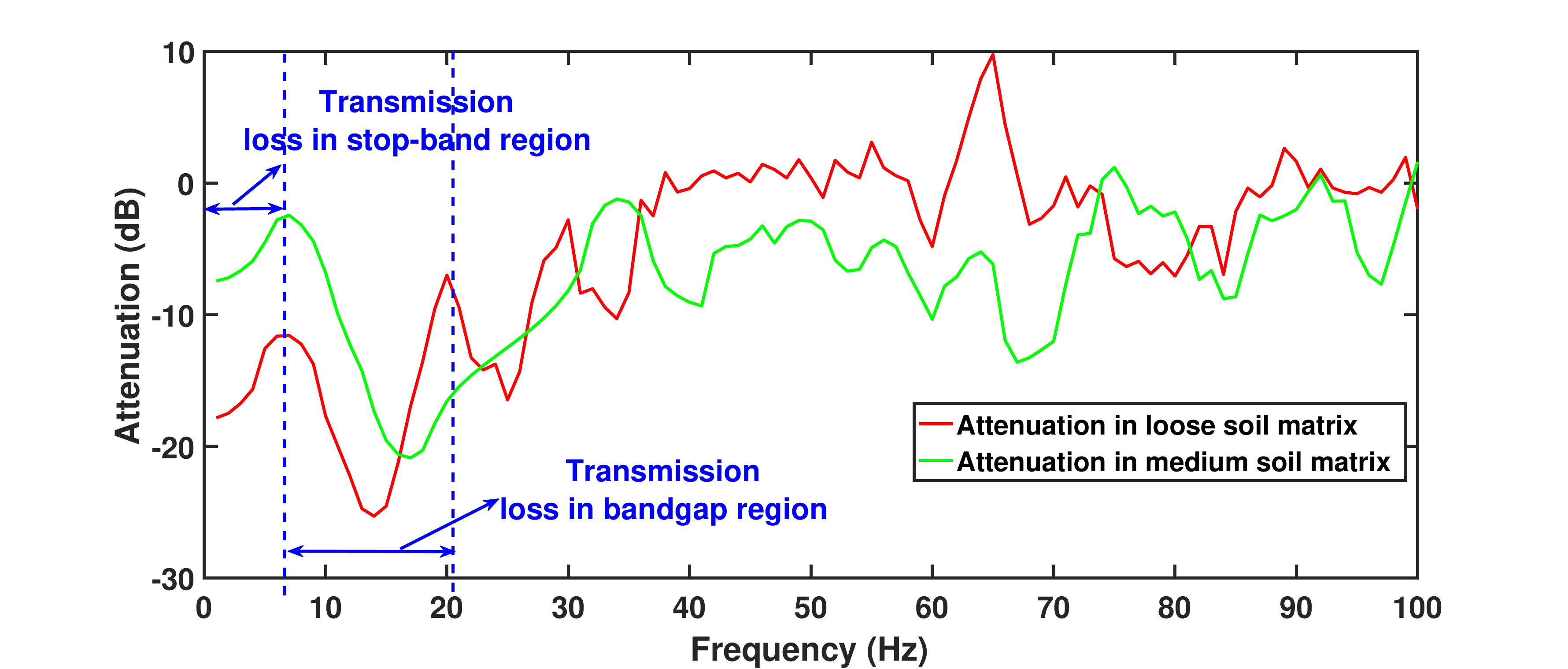}}
		{\caption{\label{F10} Attenuation for clamped configuration of square concrete columns having dimension 2 m $\times$ 2 m $\times$ 10 m in soft soil type 1 ($\text{E}=10 \ \text{MPa}$, $\mu=0.25$ and $\rho=1400 \  \text{kg}/\text{m}^3$) and medium soil  ($\text{E}=153 \ \text{MPa}$, $\mu=0.33$ and $\rho=1800 \  \text{kg}/\text{m}^3$) matrix having dimension 3 m $\times$ 3 m $\times$ 10 m (substitution ratio 0.445). Clearly, loss in soft soil type 1 matrix is more in comparison to medium soil.}}
	\end{center}
\end{figure}

So far we have discussed  the geometry of the inclusions within the unit cell, so as to generate a wide bandgap and its positioning in the frequency spectrum. However, we have exclusively utilized steel as the constituent material of the inclusion. But it is not feasible, either economically and technically, to generate such a large volume of steel having dimensions $2 \ \text{m}\times2 \ \text{m}\times10 \ \text{m}$ making up a volume of $40 \ \text{m}^3$ and then give them complex cross-section geometry via standard industrial fabrication processes. Conversely, concrete has an advantage over steel, in terms of fire and corrosion resistance, low density, resistant to impact of high groundwater level and other on-field applications like being easily mouldable into desired shapes, and these considerations make it a better choice within the civil engineering domain. Noting this fact, we compared the dispersion properties of concrete inclusions (M50 grade) with that of steel inclusions (Figs. \ref{F1d} and \ref{F4b}); both having same substitution ratio (0.445) and embedded in medium soil whose geometry and elastic properties are same as in section \ref{S31}. The elastic properties of  M50 grade concrete is taken as $\text{E}=35.35 \ \text{GPa}$, Poisson's ratio, $\mu=0.15$ and density, $\rho=2400 \  \text{kg}/\text{m}^3$. Unfortunately,  concrete inclusions in medium soil matrix show no bandgap frequencies (Fig. \ref{F4b}). However, as discussed earlier, concrete is an ideal choice for most of the civil installations, and so we performed a parametric study with different grades of concrete combined with soft, medium and dense soil. In the process of which, we arrive at a design of SM having M50 grade concrete inclusion in very soft soil type 1 (soil properties is taken as $\text{E}=10 \ \text{MPa}$, Poisson's ratio, $\mu=0.25$ and density, $\rho=1400 \  \text{kg}/\text{m}^3$ ), showing a bandgap of approximately 3.91 Hz in the range of 18.45 Hz-22.36 Hz (Fig. \ref{F4d}). Transmission loss are computed using a finite strip of five concrete square columns in very soft soil type 1 which shows large Rayleigh wave attenuation in range of 15 Hz- 25 Hz, in comparison to medium soil (Fig. \ref{F5}). Zero frequency stop-band computation (see Fig. \ref{F6}) distinctly shows two bandgaps; ultra-low stop-band frequency (0-7.62 Hz) and a higher range of bandgap, approximately 3.28 Hz (range 18.72 Hz-22 Hz). Transmission loss for $0\degree$ oriented regular-shaped square concrete columns in very soft soil type 1 ($\text{E}=10 \ \text{MPa}$, $\mu=0.25$ and $\rho=1400 \  \text{kg}/\text{m}^3$) and medium ($\text{E}=153 \ \text{MPa}$, $\mu=0.3$ and $\rho=1800 \  \text{kg}/\text{m}^3$) soil matrix is also computed over a finite array of the design region (see Fig. \ref{F10}). Clearly, concrete columns in soft soil type 1 attenuates the Rayleigh wave vibrations to a larger extent in comparison to medium soil matrix in stop-band.

\begin{figure}[H]
	\begin{center}
		{\includegraphics[width=0.8\textwidth]{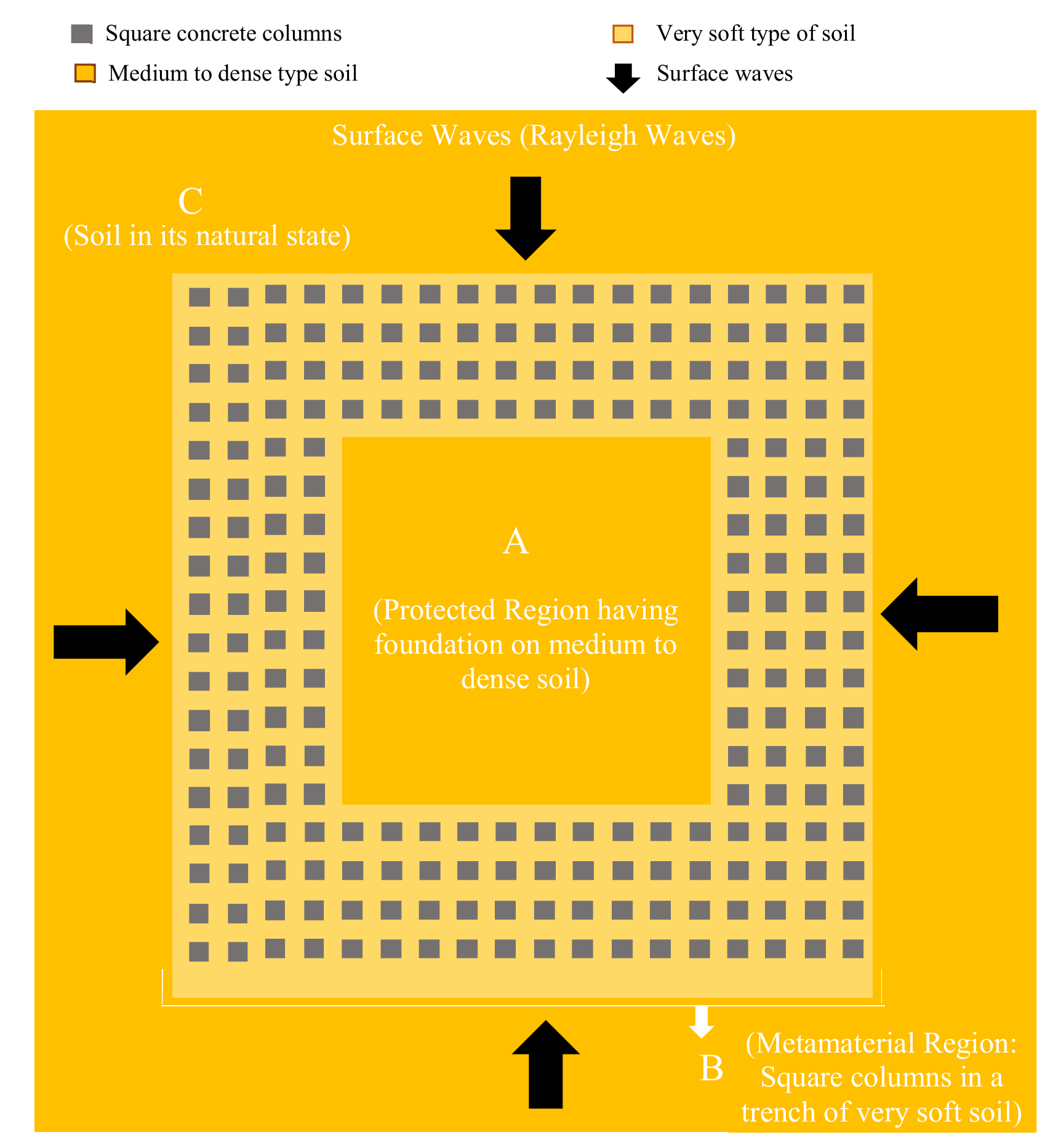}}
		{\caption{\label{F7}Schematic representation of the proposed design for metamaterial. Square columns of M50 grade embedded in loose soil matrix with properties, $\text{E}=10 \ \text{MPa}$, $\mu=0.25$ and $\rho=1400 \  \text{kg}/\text{m}^3$ are placed along a trench surrounding the protected area. The square concrete columns have substitution ratio 0.445 in the unit cell and are not clamped to the bottom.}}
	\end{center}
\end{figure}

\begin{figure}[H]
	\begin{center}
		{\includegraphics[width=0.9\textwidth]{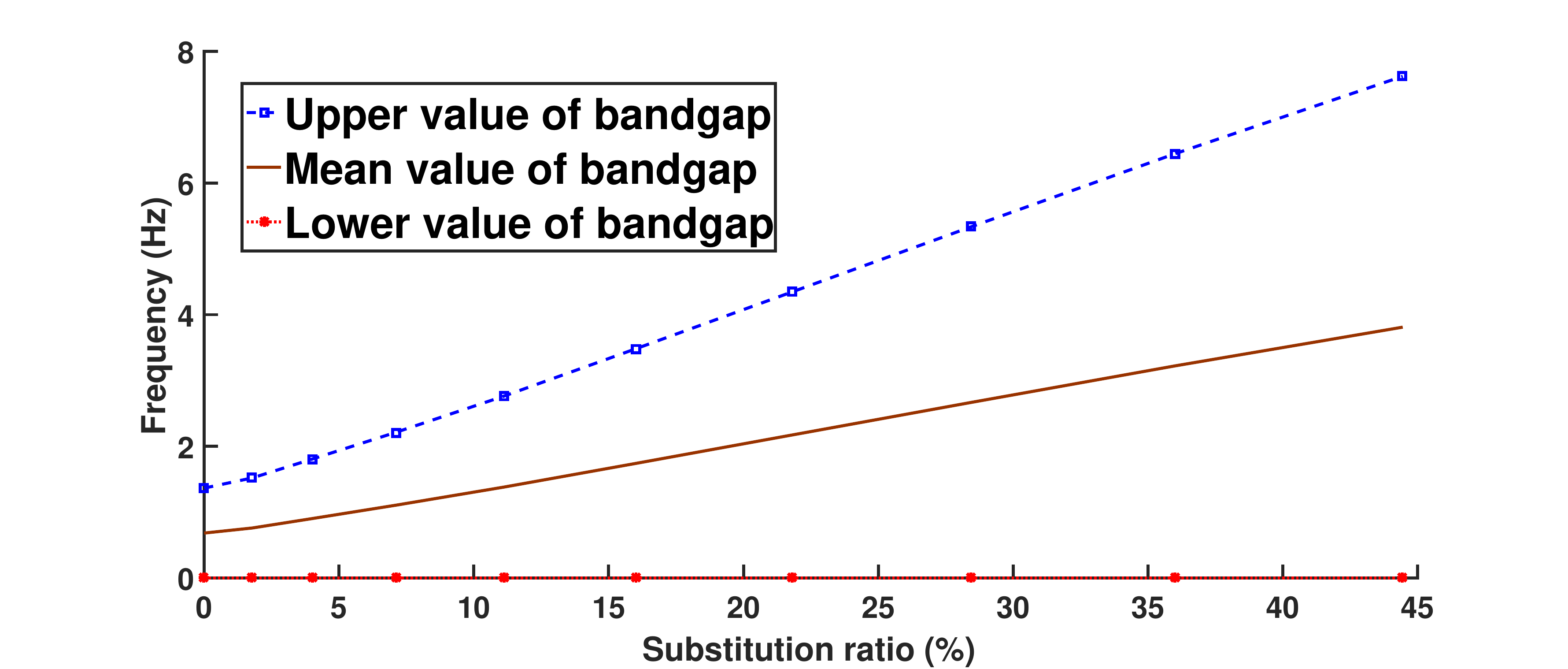}}
		{\caption{\label{F11} Bandgap versus substitution ratio plot for clamped configuration of square concrete inclusion in a unit cell of dimension 3 m $\times$ 3 m. The soil properties are taken as $\text{E}=10 \ \text{MPa}$, $\mu=0.25$ and $\rho=1400 \  \text{kg}/\text{m}^3$. }}
	\end{center}
\end{figure}

As mentioned earlier, since loose soil is not suitable for construction purpose, we recommend the following architecture. As shown in Fig. \ref{F7}, for a finite region to be protected (denoted with A in Fig. \ref{F7}), the SM microstructure should be designed using square inclusions of concrete columns oriented at $0\degree$ in a narrow strip of very soft soil matrix denoted with orange color in Fig. \ref{F7}, leaving the residual region in its natural/structured state, making civil constructions possible on the residual region. Another aspect is that since the civil installations within the protected region may vary largely, with varied natural frequency range and importance, concrete columns of such a high dimension may not be necessary always. For the same, a parametric study on obtaining stop-bands with different concrete sizes is performed and the results are shown in Fig. \ref{F11}. The effect of variation of stop-band follows a linear trend with  increasing substitution ratio of regular-shaped square concrete columns. Such graphs can be used for calibration purpose for optimal SM design.

\section{Conclusion}
\label{S4}
In this paper, we have explored the effect of different microstructure geometries of SMs towards obtaining a wider bandgap with equal amount of inclusions and its positioning in frequency spectrum, i.e., a lower bandgap region (approx. below 20 Hz) is essential for most of the engineering applications. We have arrived at a configuration of columns having a square cross-section with sides aligned with the square lattice, that gives a wider bandgap in comparison to circular cross-section inclusions for same substitution ratio of constituent material. This has been demonstrated by computing transmission losses in addition to the dispersion curves for different microstructure geometries of the SM. We have also performed analysis of the constituent material, observing that, not only steel, but concrete the conventional construction material, can also be used to design inclusions for large wave attenuation. This is important for the industrial scale-up of the technology because, concrete is cost effective, easy to cast directly at the construction site and easy to provide arbitrary geometry of the microstructure. We have observed that concrete shows a bandgap for very soft soil. Since, very soft soil is not suitable for constructions, we prescribe an architecture of SM, with concrete inclusions in a narrow strip of very soft soil matrix surrounding the protected region. Even though we have arrived at an efficient design of SM, our study open the door to further generalisation. For example, we assume linear elastic materials and it would be interesting to see what happens in case of material and geometric nonlinearity; soil often shows a significant amount of plastic deformation. In a future study, we would like to pay more attention to damping of ground vibrations, introducing so-called K-Dampers such as used to reduce vibration of bridges \cite{sapountzakis2018} and to mitigation of site-city interactions \cite{ungureanu2019influence} using the SMs which we introduce here. 

\section*{Acknowledgement}

Saikat Sarkar acknowledges Ravi Sharma and Anshul Srivastava (IIT Indore, B.Tech batch 2020) for helping him in some initial simulations at the preliminary stage of this work.


\bibliography{seismic_metamaterial_p1.bbl}

\appendix

\section{Material Properties}
For convenience we summarize the geometry and elastic properties of the microstructures in the following tables.

\begin{table}[H]
    \caption{Geometry of the microstructures}
	\label{Result17}
	 \resizebox{1\textwidth}{!}{%
	\begin{tabular}{|l|c|c|c|c|}
		\hline
		\multicolumn{1}{|c|}{\textbf{Types}} & \textbf{Width/thickness (m)} & \textbf{Length (m)} & \textbf{Diameter (m)} & \textbf{Substitution ratio} \\ \hline
		Medium soil & 3 & 3 & - & - \\ \hline
		Loose soil & 3 & 3 & - & -\\ \hline
		Circular inclusion & - & - & 1.128 & 0.445 \\ \hline
		\begin{tabular}[c]{@{}l@{}}Regular shaped\\  sqaure inclusion\end{tabular} & 2 & 2 & - & 0.445 \\ \hline
		\begin{tabular}[c]{@{}l@{}}Notch shaped\\  square inclusion\end{tabular} & 2.25 & 2.25 & - & 0.445\\ \hline
		\begin{tabular}[c]{@{}l@{}}Labyrithine\\  like inclusion\end{tabular} & 0.2 & 7 & - & 0.445\\ \hline
		\begin{tabular}[c]{@{}l@{}}4 gap split ring\\  inclusion\end{tabular} & - & - & \begin{tabular}[c]{@{}c@{}}Outer 1.28 \\ Inner 0.6\end{tabular} & 0.445 \\ \hline
		\begin{tabular}[c]{@{}l@{}}2 gap split ring \\ inclusion\end{tabular} & - & - & \begin{tabular}[c]{@{}c@{}}Outer 1.383\\ Inner 0.8\end{tabular} & 0.445 \\ \hline
		Swiss roll 1 & 0.128 & - & 1.128 & 0.445 \\ \hline
		Swiss roll 2 & 0.03 & - & 1.128 & 0.15 \\ \hline
		\begin{tabular}[c]{@{}l@{}}Seismic resonant\\ inclusion\end{tabular} & - & - & 0 - 1.311  & 0 - 0.6 \\ \hline
	\end{tabular}
}
\end{table}

\begin{table}[H]
	\caption{Material properties}
	\resizebox{1\textwidth}{!}{%
	\begin{tabular}{|l|c|c|c|}
		\hline
		\multicolumn{1}{|c|}{\textbf{Material}} & \textbf{Elastic modulus (MPa)} & \textbf{Poisson's ratio} & \textbf{Density (kg/m3)} \\ \hline
		Medium soil & 153 & 0.3 & 1800 \\ \hline
		Soft soil & 96.5 & 0.33 & 1650 \\ \hline
		Very soft clay type 1 & 10 & 0.25 & 1400 \\ \hline
		Very soft clay type 2 &  5 & 0.35 & 1633 \\ \hline
		Steel & 200.000 & 0.33 & 7850 \\ \hline
		Concrete & 35.350 & 0.15 & 2400 \\ \hline
	\end{tabular}
}
\end{table}

\begin{figure}[H] 
	\begin{subfigure}[b]{0.3\linewidth}
	\centering
		\includegraphics[width=1.5\linewidth]{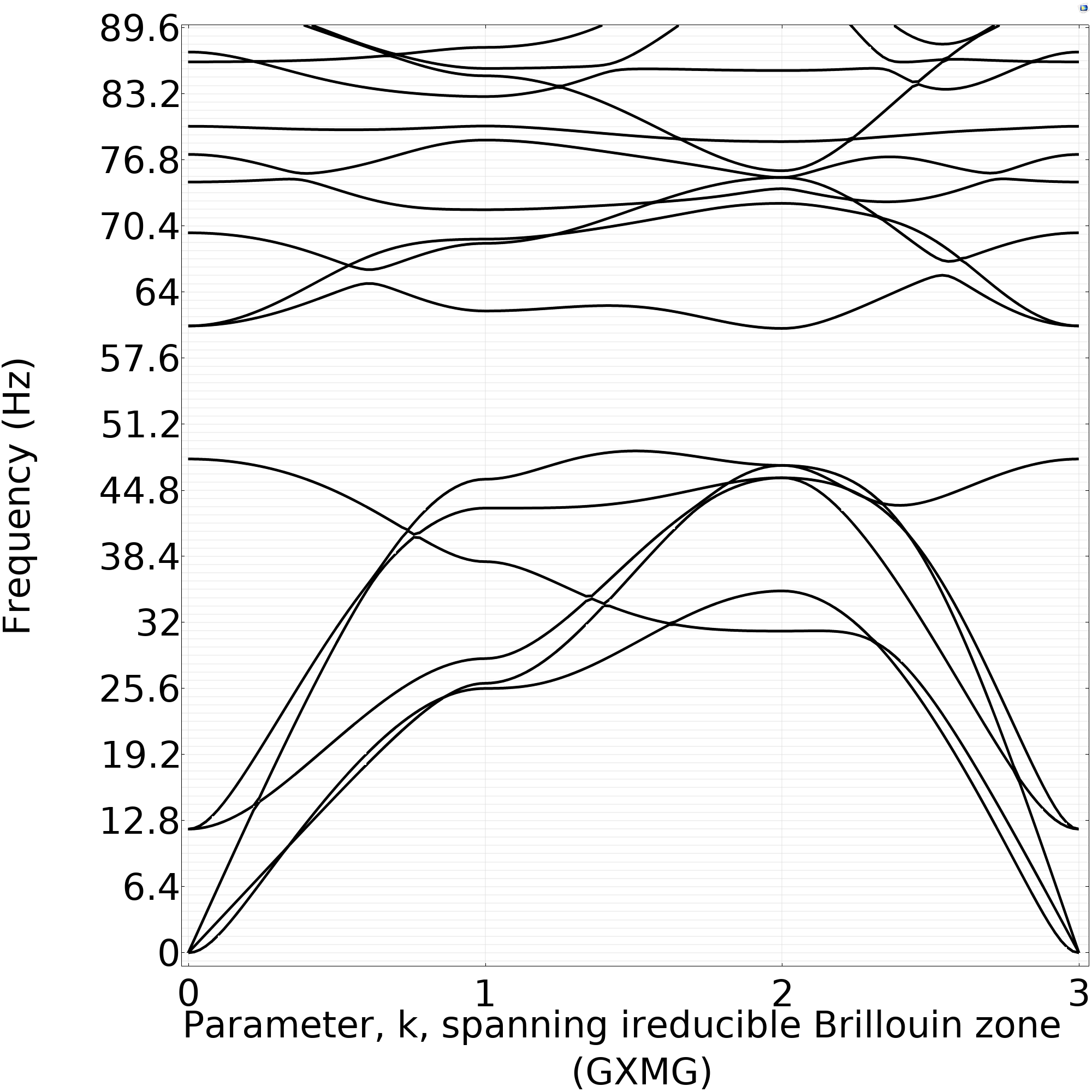} 
		\caption{} 
		\label{F33a} 
	\end{subfigure}
	\hspace{2cm}
	\begin{subfigure}[b]{0.3\textwidth}
		\centering
		\includegraphics[width=1.5\linewidth]{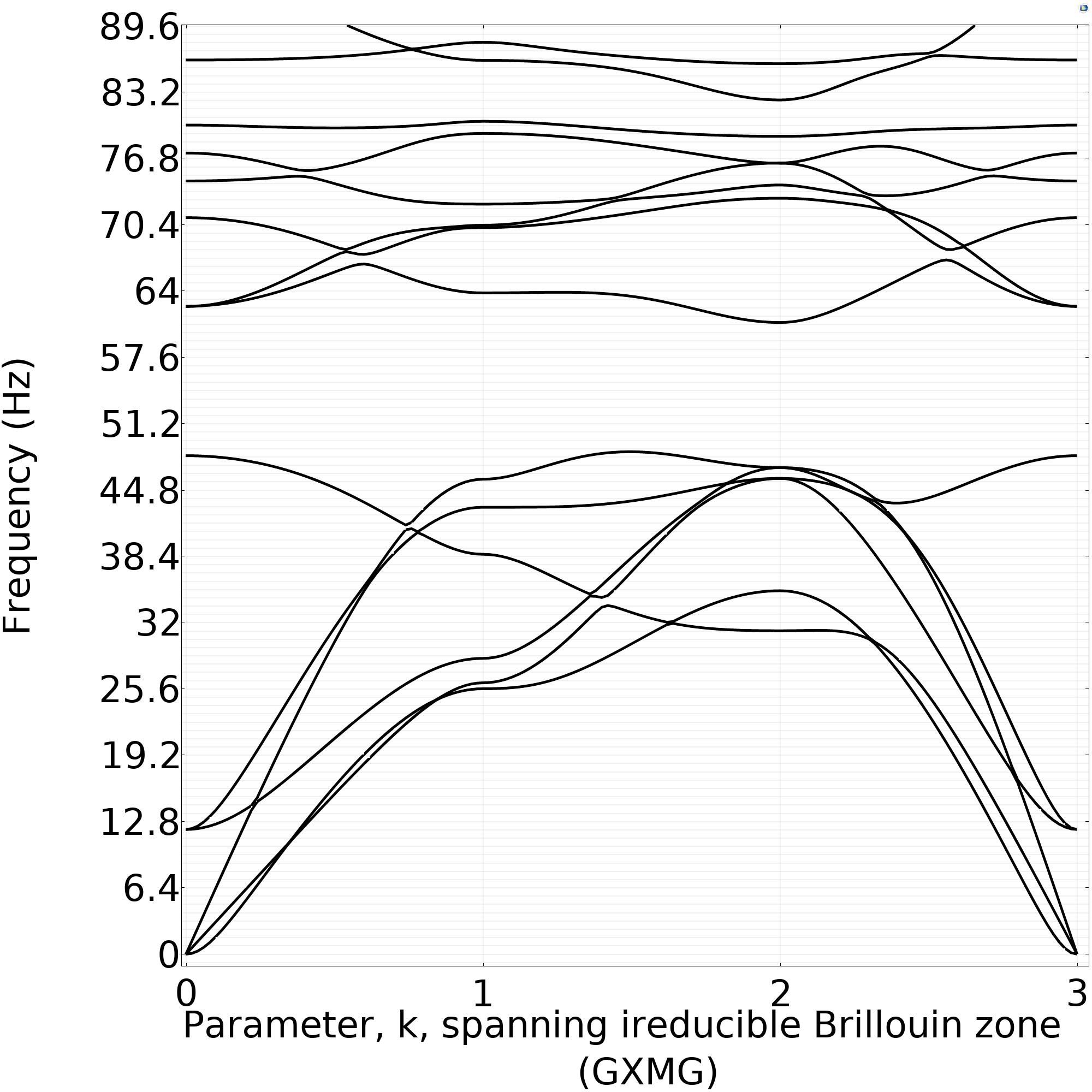} 
		\caption{} 
		\label{F33b} 
		
	\end{subfigure} 
	\caption{The Floquet-Bloch band diagrams for cylindrical inclusions are obtained around the edges of the irreducible Brillouin zone $\Gamma X M$. Note that there is an enhanced bandgap range with twisted 'ligaments' inclusions (a) in comparison to cylindrical inclusion with straight ligaments (b): the bandgap width is approximatively 12 Hz for the former and 15 Hz for the latter, in accordance with earlier study on inertial resontaors \cite{bigoni2013}. Beside that note that a new gap is opening around 80 Hz.}
	\label{F33} 
\end{figure}

\end{document}